\documentclass[12pt]{article}
\pdfoutput=1
  
\newlength{\abstractwidth}
\usepackage{epsf}
\usepackage{color}
\usepackage{graphicx}
\usepackage{tikz}
\usepackage{hyperref}

\usepackage{amsmath}
\usepackage{amssymb}
\usepackage{latexsym}

\renewcommand{\thefootnote}{\fnsymbol{footnote}}
\renewcommand{\thanks}[1]{\footnote{#1}}
\newcommand{\starttext}{
\setcounter{footnote}{0}
\renewcommand{\thefootnote}{\arabic{footnote}}}

\newcommand{\bea}{\begin{eqnarray}}
\newcommand{\eea}{\end{eqnarray}}
\newcommand{\be}{\begin{eqnarray}}
\newcommand{\ee}{\end{eqnarray}}

\def\cA{{\cal A}}
\def\cB{{\cal B}}

\def\cE{{\cal E}}
\def\cF{{\cal F}}

\def\cH{{\cal H}}
\def\cI{{\cal I}}

\def\cM{{\cal M}}
\def\cN{{\cal N}}
\def\cO{{\cal O}}

\def\cR{{\cal R}}
\def\cS{{\cal S}}

\def\cZ{{\cal Z}}

\def\ZZ{{\mathbb Z}}

\def\Re{{\rm Re \,}}
\def\Im{{\rm Im \,}}

\def\det{{\rm det \,}}

\def\eps{\epsilon}

\newcommand{\de}{{\rm d}}
\newcommand{\I}{{\rm i}}
\newcommand{\nn}{\nonumber}
\newcommand{\IZ}{\mathbb{Z}}
\newcommand{\IR}{\mathbb{R}}

\newcommand{\ttau}{S}
\newcommand{\rrho}{\tau}

\begin{document}
\starttext
\setcounter{footnote}{0}

\begin{flushright}
arXiv:1810.11343v2
\end{flushright}

\vskip 0.3in

\begin{center}

{\Large \bf String theory integrands and supergravity divergences}

\vskip 0.2in

{\large  Boris Pioline}

\vskip 0.15in

{\sl  
Laboratoire de Physique Th\'eorique et Hautes \'Energies (LPTHE), \\
CNRS and Sorbonne Universit\'e, UMR 7589, \\ 
4 place Jussieu, F-75005 Paris, France
}

\vskip 0.15in

{\tt \small pioline@lpthe.jussieu.fr}

\vskip 0.2in
October 26, 2018

\vskip 0.2in

\begin{abstract}
\vskip 0.1in
At low energies, interactions of massless particles in type II strings compactified on a torus $T^d$ are described by an effective Wilsonian action $\mathcal{S}(\Lambda)$, consisting of the usual supergravity Lagrangian
supplemented by an infinite series of higher-derivative vertices, including the much studied 
$\nabla^{4p+6q} \mathcal{R}^4$ gravitational interactions.
Using recent  results on the asymptotics of the integrands governing four-graviton
scattering at genus one and  two, I 
determine the $\Lambda$-dependence of the coefficient of the 
above interaction, and show that the logarithmic terms appearing in the  limit $\Lambda\to 0$ are related to UV divergences in supergravity amplitudes, augmented by stringy interactions. This provides a strong consistency check on the expansion of the integrand
near the boundaries of moduli space, in particular it elucidates the appearance of odd zeta values 
in these expansions. I briefly discuss how these logarithms are reflected in non-analytic
terms in the low energy expansion of the string scattering amplitude.

\end{abstract}
\end{center}

\newpage
 
\setcounter{tocdepth}{2} 
\tableofcontents


\baselineskip=15pt
\numberwithin{equation}{section}

\section{Introduction}

String theory is a manifestly UV finite theory of perturbative quantum gravity. Indeed, the $h$-th term in the weak coupling expansion of a closed string amplitude with $n$ external legs  involves an integral over the moduli space $\cM_{h,n}$ of Riemann surfaces of genus $h$ with $n$ 
punctures \cite{D'Hoker:1988ta} -- or rather, the 
moduli space $\mathfrak{M}_{h,n}$ of super-Riemann surfaces, which projects (albeit non-holomorphically) onto $\cM_{h,n}$ \cite{Witten:2012bh,Donagi:2013dua}. The latter is non-compact but its only boundaries correspond to low energy processes where the string propagates for a very long proper time. As a result, the analytic properties of scattering amplitudes of massless excitations 
around string vacua with extended supersymmetry are identical to those computed in supergravity,
supplemented with an infinite series of higher-derivative interactions induced by the exchange of massive string modes \cite{Green:1982sw}. In particular, string amplitudes around vacua of the form $\IR^{1,D-1} \times X_d$, where $X_d$ is a compact manifold of dimension $d=10-D$, are IR finite when $D>4$, and have the standard infrared divergences of supergravity in $D\leq 4$. In either case, they are non-analytic
functions of the Mandelstam variables $s_{ij} = -2 k_i\cdot k_j$ in the low energy limit $s_{ij}\to 0$, due to the exchange of massless supergravity modes. 

In order to separate these infrared singularities from genuinely stringy effects, it is convenient to decompose the moduli space $\cM_{h,n}$ into a disjoint union $\cM_{h,n}(\Lambda)\sqcup \cN_{h,n}(\Lambda)$, where 
$\cM_{h,n}(\Lambda)$ is the subset of $\cM_{h,n}$ where the proper time $t$ associated to any handle
 is less than $L=1/(\alpha'\Lambda^2)$,
 and $\cN_{h,n}(\Lambda)$ is its complement. Here, $\Lambda$ is an arbitrary intermediate scale, much larger than the energy of the external particles but much smaller than the string scale $1/\sqrt{\alpha'}$. The integral over the compact component $\cM_{h,n}(\Lambda)$ is now analytic as $s_{ij}\to 0$, and the resulting contribution to the amplitude (which we refer to as the truncated string amplitude) can be described by local interactions in an effective Wilsonian action $\cS(\Lambda)$ for the supergravity modes (see \cite{Sen:2016qap}
for an alternative definition of the  Wilsonian action in string field theory), which reduces to
the 1-particle irreducible (1PI) effective action as $\Lambda\to 0$. The integral over the non-compact complement $\cN_{h,n}(\Lambda)$ can be interpreted as a 
sum of quantum field theory amplitudes with $\ell<h$ loops computed from the action $\cS(\Lambda)$, with a UV cut-off $t>L$ on the proper time associated to each loop (and possibly an IR cut-off $t<1/(\alpha'\mu^2)$ when $D\leq 4$). The sum of the two contributions is by construction independent of $L$, with the variation of $L$ defining a renormalization group flow on the coefficients of the local couplings in the action $\cS(\Lambda)$. Thus, the dependence of  the truncated string amplitude on the
sliding scale $\Lambda$ (which acts as an IR cut-off in string theory) contains information about the dependence of the field theory amplitudes on the UV cut-off. The former is in turn governed by the behavior of the integrand of the string amplitude near the boundaries of $\cM_{h,n}$.
 
The goal of this note will be to demonstrate this connection explicitly in the case of the four-graviton scattering amplitudes in type II strings compactified on a torus $T^d$, extending the
earlier analysis of \cite{Green:2010sp} to higher order in the derivative expansion. 
In the limit where the Mandelstam variables $s=s_{12},t=s_{13},u=s_{14}$ go to zero, this amplitude can be expanded as \cite{Green:1999pv}
\be
\label{Adecomp}
\cA(s,t,u) = \cA_{\rm non.an.}(s,t,u;L) + \sum_{p,q=0}^{\infty} \cE_{(p,q)}^{(d)}(\varphi,L) \, \frac{\sigma_2^p \sigma_3^q}{p! \,q!}\, \cR^4 \ell_D^8
\ee
where $\sigma_n=(s^n+t^n+u^n) \ell_D ^{2n}/4^n$ with $n=2,3$ form a basis of symmetric functions
of $s,t,u$ subject to the momentum conservation condition $\sigma_1=0$. Here, $\ell_D$ is the $D$-dimensional Planck length, $\cR^4$ is a polynomial of degree 8 in momenta and degree 4 in the polarization tensor of the gravitons,  and $\cE_{(p,q)}(\varphi,L)$ are the coefficients
of local interactions of the form $\ell_D^{8-D-2p-3q} \nabla^{4p+6q}\cR^4$ in the Wilsonian action 
$\cS(\Lambda)$ in Einstein frame. These coefficients are functions of the moduli fields $\varphi$ 
parametrizing the constant metric and gauge forms on the internal torus as well as the $D$-dimensional string coupling $g_D$. In maximally supersymmetric vacua, $\varphi$ is valued 
in the symmetric space  
$\cM_D=E_{d+1}(\IZ) \backslash E_{d+1}/ K_{d+1}$, where $E_{d+1}$ is the split reductive 
Lie group of Cartan type $E_{d+1}$, $K_{d+1}$ its maximal compact subgroup and $E_{d+1}(\IZ)$
an arithmetic subgroup of $E_{d+1}$ known as the U-duality group \cite{Hull:1994ys}. 

The first term $\cA_{\rm non.an.}(s,t,u;L)$ in \eqref{Adecomp}
corresponds to the integral over the domains 
$\cN_{h,n}(\Lambda)$, and can be viewed as a supergravity amplitude with UV cut-off 
$\Lambda=1/\sqrt{\alpha' L}$, involving the usual massless propagators and supergravity vertices, supplemented with higher derivative interactions.  In the limit $L\to\infty$, the coefficients 
$\cE_{(p,q)}^{(d)}(\varphi,L)$ are typically divergent, but these divergences are cancelled by
$L$-dependent contributions from $\cA_{\rm non.an.}(s,t,u;L)$ 
proportional to $\sigma_2^p \sigma_3^q\, \cR^4$.
After subtracting these divergent terms, one obtains finite, `renormalized' 
automorphic functions $\cE_{(p,q)}^{(d)}(\varphi)$ on $\cM_D$, which have been intensively studied 
in recent years (see e.g. \cite{Pioline:2015yea} for a recent entry point in this vast literature). 
We stress that the adjective `renormalized' is {\it not} related to the usual renormalization prescription in quantum field theory, but instead
refers to a specific prescription for separating the analytic and non-analytic part of the 1PI effective
action.

As indicated earlier, the coefficient $\cE_{(p,q)}^{(d)}(\varphi,L)$ of the $\nabla^{4p+6q}\cR^4$
effective interaction is given, up to {\it a priori unknown} non-perturbative effects, by  a
genus expansion
\be
\cE_{(p,q)}^{(d)}(\varphi,L) = g_D^{\frac{2D-8p-12q-16}{D-2}} \sum_{h=0}^{\infty} 
g_D^{2h-2}\, \cE_{(p,q)}^{(d,h)}(\varphi,L) + \cO(e^{-2\pi /g_D})\ ,
\ee
where $\cE_{(p,q)}^{(d,h)}(\varphi,L)$ is given by an integral over the truncated moduli space
$\cM_{h,4}(\Lambda)$, and the prefactor arises when translating from string frame to Einstein frame
(or equivalently, when expressing the string length $\ell_s$ in terms of the Planck length $\ell_D=\ell_s\, g_D^{2/(D-2)}$).
The renormalized coupling $\cE_{(p,q)}^{(d)}(\varphi)$ has a similar genus expansion, 
except for an additional contribution $\cE_{(p,q)}^{(d),{\rm non. an.}}$ 
which is non-analytic at $g_D=0$ and arises in the process of translating 
$\cA_{\rm non.an.}(s,t,u;L)$ into Einstein frame \cite{Green:2010sp,Pioline:2015yea,Basu:2016kon}.
At tree-level, the coefficients are pure constants known to arbitrary order, the first few ones 
being \cite{Gross:1986iv,Green:1999pv}
 \be
 \label{Epqtree}
 \cE_{(0,0)}^{(d,0)} = 2\zeta(3),\quad
 \cE_{(1,0)}^{(d,0)} = \zeta(5), \quad 
 \cE_{(0,1)}^{(d,0)} = \frac23 \zeta(3)^2,\quad
  \cE_{(2,0)}^{(d,0)} = \zeta(7),\quad
  \cE_{(1,1)}^{(d,0)} = \frac23 \zeta(3) \zeta(5)
  \dots
 \eea
 
At genus-one, the scattering amplitude is known since \cite{Green:1981yb} and its 
low energy  expansion has been studied to high order 
in the $\alpha'$ expansion \cite{Green:1999pv,Green:2008uj,D'Hoker:2015foa}. 
After integrating over the location of the four vertex
operators, the coefficients $\cE_{(p,q)}^{(d,1)}(\varphi,L)$ are given by integrals of modular graph functions \cite{DHoker:2015wxz} over the truncated fundamental domain $\cF_1(L)$ in the Poincar\'e upper-half plane $\cH_1$. The cancellation of the $L$-dependent terms against the
non-analytic part $\cA_{\rm non.an.}(s,t,u;L)$ was studied in detail in \cite{Green:2008uj},
primarily in dimensions $D=10$ and $D=9$, using unitarity to determine the threshold contributions.

The genus-two amplitude was computed in a heroic effort
\cite{DHoker:2001kkt,D'Hoker:2002gw,D'Hoker:2005jc} 
and its low energy expansion was investigated in 
\cite{D'Hoker:2013eea,DHoker:2014oxd,DHoker:2017pvk}. 
The coefficients $\cE_{(p,q)}^{(d,2)}(\varphi,L)$ are again given by integrals of 
genus-two modular graph functions \cite{DHoker:2017pvk} over the truncated fundamental domain $\cF_2(L)$ in the Siegel upper-half plane \cite{Pioline:2015nfa}. 
Recently,  an extensive analysis of the asymptotic behavior of the 
genus-two integrands near the separating and non-separating degeneration limits
was carried out up to order $\nabla^8\cR^4$ \cite{DHoker:2017pvk,DHoker:2018mys}, extending earlier 
work  on the supergravity limit of genus-two string amplitudes \cite{Green:2008bf}.

In Section \ref{sec_trunc}, we shall  use these asymptotics to compute the $L$-dependence of the truncated amplitude at genus-two, and extract the coefficients of the logarithmic terms in various dimensions. In Section \ref{sec_sugra}  
we compare these results to the structure of ultraviolet divergences in maximally
supersymmetric supergravity, obtained up to three loop in  \cite{Bern:1998ug,Bern:2008pv}. 
We shall find quantitative agreement, providing a strong check on the asymptotics computed in 
\cite{DHoker:2018mys}, as well as elucidating the physical origin of the transcendental  
coefficients  which appear in these expansions. In Section \ref{sec_discuss}, we briefly
discuss the implications of these logarithms for the non-analytic part of the amplitude. 
The detailed study of the cancellation
of the $L$-dependent terms against the non-analytic part of the 1PI effective action is left
for future work. While we restrict attention to interactions up to order $\nabla^{8}\cR^4$
in the body of the paper,  the constraints on the
asymptotics of the genus-two integrand at order $\nabla^{10}\cR^4$ 
are investigated in Appendix  \ref{sec_d10r4}. Finally, some details about a class of modular functions 
appearing in the tropical
limit of genus two string amplitudes are collected in  Appendix \ref{sec_local}. 

\medskip
{\it Note added in v2:} In the first release of this work, the truncated moduli space $\cM_{h,n}(\Lambda)$ was defined as the subspace of $\cM_{h,n}$ where the length of any closed geodesic measured with respect to the hyperbolic metric is less than $L=1/(\alpha'\Lambda^2)$. This definition does not make sense since closed geodesics typically have unbounded length. A more sensible definition is to put a lower cut-off $\ell>\epsilon$ on the length of the shortest closed geodesic (also known as the systole). 
Indeed, from the point of view of hyperbolic metric, all boundaries of $\cM_{h,n}$ are associated to 
the vanishing $\ell\to 0$ of the length of a closed geodesic (homologically trivial in the separating case,
or non-trivial in the non-separating case). In the vicinity of a locus where $\ell\to 0$, the geometry is 
 well approximated by the wedge 
 $\{ \frac{\ell}{2} < \arg z < \pi-\frac{\ell}{2}\}$ inside the Poincar\'e upper half-plane, modded out by dilations $z\mapsto e^\ell z$ -- a model known as the standard collar (see e.g. \cite{Wolpert85,Moosavian:2017qsp}). 
 This region is conformal
 to  the cone $xy=\tau$ with $\log|\tau|\sim -2\pi^2/\ell$. 
As the length $\ell$ of the geodesic along the imaginary axis shrinks to 0 size, the length 
$\ell'=\int_{\ell/2}^{\pi-\ell/2} \frac{\de\theta}{\sin\theta}$ of a geodesic at any fixed radial distance  diverges as $\ell'\sim -2\log \ell$. A lower cut-off $\ell>\epsilon$ therefore implies an upper bound $
\ell'<-2\log\epsilon$ on the length of geodesics going through the node. In view of the relation
 $\Omega_{ij}\sim \frac{1}{2\pi\I} \log \tau$ between the gluing parameter $\tau$ and the relevant entry  $\Omega_{ij}$ in the period matrix
(in the case of a non-separating degeneration) \cite{Fay73}, 
one readily sees that a lower cut-off $\ell>\epsilon$
implies an upper bound $t \sim \Im\Omega_{ij} < \pi/\epsilon$ on the proper time associated to the 
long  handle (for a separating degeneration, the relevant entry $\Omega_{ij}$ is proportional to $\tau$,
but exponential in proper time, leading to the same conclusion). 
Thus, we can define  $\cM_{h,n}(\Lambda)$ as the subspace of $\cM_{h,n}$ where
closed geodesics with respect to the hyperbolic metric have length greater than
$\pi/L= \pi \alpha' \Lambda^2$. It would be interesting
to give a similar characterization in terms of the minimal area metric which is more commonly used in the string field theory literature \cite{Zwiebach:1990nh}.

\section{Scale dependence of truncated modular integrals}
\setcounter{equation}{0}
 \label{sec_trunc}
In this section, we discuss the scale dependence of the modular integrals computing the
coefficient $\cE_{(p,q)}^{(d,h)}(\varphi,L)$ of the four-graviton effective interaction in type II string theory compactified on a torus $T^d$, at genus $h=1$ and $h=2$. We also briefly
review the genus three result at leading order in the derivative expansion.

\subsection{Genus-one \label{sec_amp1}}
By expanding the integrand of the genus-one four-graviton scattering in powers of the 
Mandelstam variables at fixed value of the complex modulus $\tau$ of the worldsheet torus, 
and integrating
over the position of the four vertex operators, one finds a contribution to 
the effective coupling $\cE_{(p,q)}^{(d)}$  of the form \cite{Green:2008uj}
\be
\label{Epq1reg}
\cE_{(p,q)}^{(d,1)} (\varphi,L)= \pi \int_{\cF_1(L)} \de\mu_1\, \Gamma_{d,d,1}(\tau;\varphi)\,
\cB^{(1)}_{(p,q)}(\tau)\,  ,
\ee
where $\de\mu_1=2\de\tau_1\de\tau_2/\tau_2^2$ is the  invariant measure on the
Poincar\'e upper-half plane,  
 $\Gamma_{d,d,1}$ is the genus-one Siegel-Narain theta series
 \be
 \label{Gdd1}
 \Gamma_{d,d,1}(\tau;\varphi) = \tau_2^{d/2}\, \sum_{(m_i, n^i)\in\IZ^{2d}}
 e^{-\pi \tau_2[ (m_i +B_{ij n^j} G^{ik}(m_k+B_{kl} n^l)+n^i G_{ij} n^j] + 2\pi\I m_i n^i \tau_1}\ ,
 \ee
 where $G_{ij}$ is the constant metric on the torus $T^d$, $G^{ij}$ its inverse and
 $B_{ij}$ the Kalb-Ramond two-form -- all being functions of the moduli $\varphi$. 
 The factor $\cB^{(1)}_{(p,q)}$ in the integrand is a non-holomorphic modular function of $\tau$, given
for low values of $(p,q)$ by  \cite{Green:2008uj} \cite[(7.17),(7.28)]{D'Hoker:2015foa}\footnote{
The factor of $p! q!$ is not included in \cite{D'Hoker:2015foa}; 
 to compare with \cite{Pioline:2015yea}, note that $E_2^*=\frac12 E_2, E_3^*=E_3$.}:
 \be
 \label{Jpq}
 \begin{split}
 \cB^{(1)}_{(0,0)}=&1,\quad
 \cB^{(1)}_{(1,0)}=E_2,\quad
 \cB^{(1)}_{(0,1)}=\frac13(5E_3+\zeta(3)),\quad
 \cB^{(1)}_{(2,0)}=&\frac{1}{12}(D_4+9E_2^2+6E_4),
    \end{split}
 \ee
 where $E_a(\tau)$ are the usual non-holomorphic Eisenstein series, normalized such that
 $E_a = 2 \tau_2^a \zeta(2a)/\pi^a + \cO(\tau_2^{1-a})$, while $D_4(\tau)$ is
the  modular graph function defined in \cite[\S 5.3]{D'Hoker:2015foa}.

Finally, $\cF_1(L)$ is a truncated version of the fundamental domain for the action of $SL(2,\IZ)$ 
on the upper half plane parametrized by $\tau=\tau_1+\I\tau_2\in \cH_1$, introduced in 
the present context  in  \cite{Green:1999pv},
\be
\label{defF1L}
\cF_1(L) = \{ -\frac12 < \tau_1 < \frac12, \quad |\tau|>1, \quad \tau_2 < L \} \ ,
\ee
where $L$ is a large but finite constant. The truncation $\tau_2 < L$ ensures that
the proper time around the loop is less than $L$ (in units where the proper length of the string
is normalized to 1). 
We shall refer to integrals of modular forms over the truncated fundamental domain
as `truncated modular integrals'. Clearly, one could choose a different fundamental domain
for the action of $SL(2,\IZ)$, but then the truncation should be appropriately
modified, such that it covers the same domain as $\cF_1(L)$ on the quotient $\cH_1/SL(2,\IZ)$.

Our purpose will be to extract the dependence of the truncated modular integral \eqref{Epq1reg}
on the scale $L$ as it is taken to infinity. The simplest way is to compute the derivative with respect 
to $L$, 
\be
L\partial_L \cE_{(p,q)}^{(d,1)} (\varphi,L)= \frac{2\pi}{L} \int_{-1/2}^{1/2} \de\tau_1\,  \left[
\Gamma_{d,d,1}(\tau;\varphi)\, \cB^{(1)}_{(p,q)}(\tau)\right]_{\tau_2=L}\ .
\ee
In the limit $\tau_2=L\to\infty$, we can approximate $\Gamma_{d,d,1}(\tau;\varphi)\sim \tau_2^{d/2}$
with exponential accuracy. We shall assume that in this limit,  the
constant term of the integrand $\cB^{(1)}_{(p,q)}$ has a finite Laurent series expansion
\be
\label{B1growth}
\int_{-1/2}^{1/2} \de\tau_1\, \cB^{(1)}_{(p,q)}(\tau) = \sum_{i=1}^{\ell} c_i \tau_2^{\sigma_i} + \cO(e^{-2\pi\tau_2}) \ ,
\ee
up to exponentially suppressed corrections.
Indeed, in the cases of interest, $\cB^{(1)}_{(p,q)}$ is a modular graph function of weight\footnote{Here $w$ denotes  the transcendental weight,  while the modular weight vanishes.} $w=2p+3q$
with a Laurent expansion of the form \eqref{B1growth} with $\sigma_i\in [1-w,\dots, w]$. Under
this assumption, we find
\be
L\partial_L \cE_{(p,q)}^{(d,1)} (\varphi,L)= 2\pi \sum_{i=1}^{\ell} c_i \, L^{\frac{d}{2}+\sigma_i-1}\ .
\ee
Upon integrating, it follows that the truncated modular integral 
\eqref{Epq1reg} behaves as $L\to\infty $ as 
\be
\label{epq1reg}
e_{(p,q)}^{(d,1)}(L)  =  2\pi \left( 
\sum_{\sigma_i> 1-\frac{d}{2}}  c_i 
\frac{L^{\frac{d}{2}+\sigma_i-1}}{\frac{d}{2}+\sigma_i-1} 
+ \sum_{\sigma_i=1-\frac{d}{2}}  c_i 
\log L \right)\ ,
\ee
up to an additive constant. 
Following the terminology of \cite{MR656029}, 
we define the `renormalized'  integral $\cE_{(p,q)}^{(d,1)}$ 
by subtracting the power-like terms before taking the limit $L\to \infty$, 
\be
\label{Epq1renorm}
\cE_{(p,q)}^{(d,1)}(\varphi) = \lim_{L\to\infty} \left[ \cE_{(p,q)}^{(d,1)}(\varphi,L) - e_{(p,q)}^{(d,1)}(L) 
\right]
\ee
Clearly, terms with $\sigma_i<1-\frac{d}{2}$ could be added to 
the sum \eqref{epq1reg} without affecting \eqref{Epq1renorm}. 
Moreover, when a contribution with $\sigma_i=1-\frac{d}{2}$
is present in the sum, one may change the scale of the logarithm, i.e. replace $\log L$ by $\log \mu L$, at the expense of shifting \eqref{Epq1renorm} by an additive constant $2\pi c_i \log\mu$.
Of course, the truncated modular integral $\cE_{(p,q)}^{(d,1)}(L)$ is independent of that choice.

Using known asymptotics of 
the modular graph functions $E_a, D_4$, the expansion  
of the integrands \eqref{Jpq} at large $\tau_2$ is given by 

\bea
\cB^{(1)}_{(1,0)} &=& \frac{\pi ^2 \tau_2^2}{45}+\frac{\zeta (3)}{\pi  \tau_2} +\cO(e^{-2\pi\tau_2}) \\
\cB^{(1)}_{(0,1)} &=&\frac{2 \pi ^3 \tau_2^3}{567}+\frac{\zeta
   (3)}{3}+\frac{5 \zeta (5)}{4 \pi ^2 \tau_2^2}+\cO(e^{-2\pi\tau_2})\\
\cB^{(1)}_{(2,0)} &=&\frac{8 \pi ^4 \tau_2^4}{14175}+\frac{4 \pi  \tau_2 \zeta (3)}{45}+\frac{5 \zeta
   (5)}{6 \pi  \tau_2}+\frac{\zeta (3)^2}{2 \pi ^2 \tau_2^2}+\frac{\zeta (7)}{2 \pi
   ^3 \tau_2^3}+\cO(e^{-2\pi\tau_2})
\eea
Thus, we find that the truncated modular integrals are given,  up to 
exponentially suppressed corrections as $L\to\infty$, by\footnote{These results
agree for $d=0$ with Eq. (3.12), (3.14), (3.17) in \cite{Green:2008uj}, up to normalisation
factors. For $d=0$, the constant term $\cE_{(p,q)}^{(0,1)}$ can be computed using the lemma in 
\cite[\S A.2]{Green:2008uj}. For $d\geq 1$, differential equations for the renormalized integrals
$\cE_{(0,0)}^{(d,1)}(\varphi)$ with respect to the torus moduli were established in 
\cite{Pioline:2015nfa,Basu:2016fpd}.}
\bea
\label{E00L1}
\cE_{(0,0)}^{(d,1)}(\varphi,L) &=& 
2\pi\, \frac{L^{\frac{d-2}{2}}}{\frac{d-2}{2}} + \cE_{(0,0)}^{(d,1)}(\varphi)
\\
\label{E10L1}
\cE_{(1,0)}^{(d,1)}(\varphi,L) &=& 
\frac{2\pi^3}{45} \frac{L^{\frac{d+2}{2}}}{\frac{d+2}{2}} 
+ 2\zeta(3) \frac{L^{\frac{d-4}{2}}}{\frac{d-4}{2}}
+ \cE_{(1,0)}^{(d,1)}(\varphi)
\\
\label{E01L1}
\cE_{(0,1)}^{(d,1)}(\varphi,L) &= &
\frac{4\pi^4}{567} \frac{L^{\frac{d+4}{2}}}{\frac{d+4}{2}} 
+\frac{2\pi}{3} \zeta(3) \frac{L^{\frac{d-2}{2}}}{\frac{d-2}{2}}  
+ \frac{5}{2\pi}\zeta(5) \frac{L^{\frac{d-6}{2}}}{\frac{d-6}{2}}
+ \cE_{(0,1)}^{(d,1)}(\varphi)
\\
\cE_{(2,0)}^{(d,1)}(\varphi,L) &=& 
 \frac{16\pi^5}{14175} \frac{L^{\frac{d+6}{2}}}{\frac{d+6}{2}} 
+\frac{8 \pi^2\zeta(3)}{45} \frac{L^{\frac{d}{2}}}{\frac{d}{2}}
+\frac{5\zeta(5)}{3}  \frac{L^{\frac{d-4}{2}}}{\frac{d-4}{2}}
+\frac{\zeta(3)^2}{\pi} \frac{L^{\frac{d-6}{2}}}{\frac{d-6}{2}}
\nn\\
&&+\frac{\zeta(7)}{\pi^2} \frac{L^{\frac{d-8}{2}}}{\frac{d-8}{2}}
+ \cE_{(2,0)}^{(d,1)}(\varphi)
\label{E20L1}
\eea
In these expressions, for a given dimension $d$,  terms proportional to $L^\alpha$ with $\alpha<0$
should be omitted, while terms of the form  $L^{\frac{d-k}{2}}/\frac{d-k}{2}$ for $d=k$ should be understood as $\log L$, as prescribed in  \eqref{epq1reg}. As we shall explain in Section \ref{sec_sugra}, the fact that the 
coefficients in these expansions involve the same odd zeta values which appear in the tree-level
couplings \eqref{Epqtree} is not a concidence.

We note that a different way of regularizing modular integrals of the form  \eqref{Epq1reg} is to  insert a power $\tau_2^\epsilon$ in the integrand,
and take the constant term in the Laurent expansion around $\epsilon=0$. This 
mimicks the effect of dimensional regularisation prescription in quantum field theory, where one is
instructed to compute the loop integral in $D-2\epsilon$ non-compact dimensions, and
take the limit $\epsilon\to 0$ after subtracting all polar terms. Note that the insertion
of $\tau_2^\epsilon$ breaks modular invariance (but not T-duality), and is only meaningful
provided $\cF_1$ is the standard `keyhole' fundamental domain, as opposed to one of its
modular transforms. This prescription is equivalent for the purpose of computing the 
coefficient of the logarithmic terms, though the scale of the log would be in 
general different (see \cite{bringmann2016regularized} for a recent mathematical discussion 
of regularized modular integrals).

\subsection{Genus-two  \label{sec_amp2}}

Similarly, by expanding the integrand of the genus-two four-graviton scattering amplitude
computed in \cite{D'Hoker:2005jc}
 in powers of Mandelstam variables at fixed values of the period matrix $\Omega$
 in the Siegel upper-half plane
 $\cH_2$, 
 one finds a contribution to the 
effective coupling $\cE_{(p,q)}^{(d)}$  of the form \cite{D'Hoker:2013eea,Pioline:2015nfa}
\be
\label{d8rfourreg}
\cE^{(d,2)}_{(p,q)}(\varphi,L) = \frac{\pi}{4} 
\int_{\cF_2(L)} \de\mu_2 \, \Gamma_{d,d,2}(\Omega;\varphi)\,\cB^{(2)}_{(p,q)}(\Omega) , 
\ee
where 
 $\de\mu_2$ is the  invariant measure on the Siegel upper-half plane $\cH_2$,  $\Gamma_{d,d,2}(\Omega;\varphi)$ is the genus-two Siegel-Narain theta series
\be
 \label{Gdd2}
\Gamma_{d,d,2}(\Omega;\varphi) = |\Omega_2|^{d/2} \sum_{(m_i^I, n^{i,I})\in \ZZ^{2d}}
e^{-\pi \mathcal{L}^{IJ} \Omega_{2,IJ} + 2\pi\I m_i^I n^{i,J} \Omega_{1,IJ}}
\ee
where, similarly to \eqref{Gdd1},
\be
\mathcal{L}^{IJ}  = (m_i^I + B_{ij} n^{j,I}) G^{ik} (m_k^J + B_{kl} n^{l,J})
+ n^{i,I} G_{ij} n^{j,J}\ .
\ee
The factor $\cB^{(2)}_{(p,q)}$ in the integrand is in general a real-analytic Siegel modular function, regular
away from the divisor $v=0$ and its images under $Sp(4,\IZ)$, and belongs to the class of genus-two
modular graph functions of weight $w=2p+3q-2$ considered in \cite{DHoker:2017pvk}. 
For low values of $(p,q)$, it is given explicitly by 
\be
\label{Bpq}
\cB^{(2)}_{(0,0)}=0\ ,\quad \cB^{(2)}_{(1,0)}=2\ ,\quad \cB^{(2)}_{(0,1)}=4\varphi_{\rm KZ}\ ,\quad 
\cB^{(2)}_{(2,0)}=\frac12(\cZ_1 -2 \cZ_2 + \cZ_3)\ ,\quad \dots
\ee
where $\varphi_{\rm KZ}$ is the Kawazumi-Zhang invariant \cite{D'Hoker:2013eea,Pioline:2015qha}, introduced
in the mathematics literature in \cite{Kawazumi,Zhang}, and $\cZ_a$, $a=1,2,3$ are the 
genus-two string invariants defined in  \cite[(2.12)]{DHoker:2018mys} 
and studied extensively there.  Finally,
\be
\label{defF2L}
\cF_2(L) = \cF_2 \cap \left\{ 
\Omega={\scriptsize \begin{pmatrix} \tau & v \\ v & \sigma \end{pmatrix}} : \quad \frac{\det\Im\Omega}{\Im\tau} < L\right\}
\ee
 is a truncated version \cite{Pioline:2015nfa} of the standard
  fundamental domain $\cF_2$ for the action of $Sp(4,\IZ)$ on $\cH_2$, whose interior
  is defined
by the conditions \cite{zbMATH03144647} \cite[A.15]{D'Hoker:2014gfa}
 \bea
 -\frac12 < \tau_1, v_1, \sigma_1 <\frac12\ , \qquad \qquad
 0 < 2 v_2 < \tau_2 <\sigma_2\ , 
  \label{fundom2}\\
 |\det(C\Omega+D)|>1\ , \mbox{for all}\ \begin{pmatrix} A & B\\C& D \end{pmatrix}\in Sp(4,\IZ)\ .
 \nn
 \eea
 To justify why \eqref{defF2L} is an appropriate regularization,  we first recall some basic
facts about the moduli space  of genus-two Riemann surfaces.


\subsubsection*{Boundaries of the moduli space  of genus-two Riemann surfaces}

The moduli space $\cM_{2}$ of genus-two Riemann surfaces without punctures is isomorphic to the complement of the divisor $\Omega_{12}=0$ in the fundamental domain $\cF_2$. Its compactification, known as the Deligne-Mumford moduli space, is obtained by adding two divisors, corresponding to the separating
and non-separating degenerations.  In order
to discuss these degenerations, it is useful to introduce 
several (redundant) coordinate systems for $\Im\Omega$,
\be
\label{defVttau}
\Im\Omega = \begin{pmatrix} L_1 + L_2 & L_1 \\ L_1 & L_1+L_3 \end{pmatrix} 
= \begin{pmatrix}\tau_2 & \tau_2 u_2 \\ \tau_2 u_2 & t+ \tau_2 u_2^2 \end{pmatrix} 
= \frac{1}{V\ttau_2}  \begin{pmatrix}1 & \ttau_1 \\ \ttau_1 &|\ttau|^2 \end{pmatrix} 
\ee
where $L_1,L_2,L_3,t,V$ are real positive variables, 
$v=u_1+\tau u_2 \in \mathbb{C}/(\IZ+\tau\IZ)$, and $\ttau=\ttau_1+\I\ttau_2$ lies in the
Poincar\'e upper half-plane. In the standard fundamental domain \eqref{fundom2}, one
has $L_1< L_2 < L_3$ while $\ttau$ lies in the positive half $\cF_1 \cap \{ \ttau_1>0\}$
of the standard fundamental domain $\cF_1$, which we denote by $\cF$. This
half-domain is in fact a fundamental domain for the action \eqref{GL2act} of $GL(2,\IZ)$  on $\cH_1$. 
In terms of these variables, the relevant degenerations are as follows\footnote{Further degenerations combining separating and non-separating degenerations 
are allowed but will not be relevant for our purposes.}:

\begin{itemize}
\item[i)] The locus 
$v=0$ (which sits inside $\cF_2$ but on the boundary of $\cM_2$) corresponds to the separating
degeneration limit where the surface $\Sigma_2$ degenerates into a pair of elliptic curves joined by a very long tube. Fortunately, the Siegel-Narain partition function \eqref{Gdd2} is regular on this locus, while $\cB^{(2)}_{(p,q)}$ grows only as a power $(\log|v|)^w$ \cite[\S 4]{DHoker:2018mys} so that the integral converges in this region. 
\item [ii)]
The `minimal' non-separating
degeneration corresponds to setting $L_3\to\infty$ keeping $L_1,L_2$ and $\Re\Omega$ fixed; or equivalently
$t\to\infty$ keeping $\tau,v,\sigma_1$ fixed. In this limit, $\Sigma_2$
degenerates into an elliptic curve $\Sigma_1$ of modulus $\tau$, with two marked points 
at $z=0$ and $z=v$ on $\Sigma_1$ (up to an irrelevant translation), and joined by a very long
tube of proper length $t$. This region is responsible for one-loop subdivergences.

\item[iii)] The `double' non-separating
degeneration corresponds to setting $L_2, L_3\to\infty$ at the same rate, keeping $L_1$ and $\Re\Omega$ fixed; equivalently
$\tau_2,\sigma_2\to\infty$ keeping $v$ fixed, or $V, S_1 \to 0$ keeping $S_1/V$ fixed. In this limit, $\Sigma_2$
degenerates into an sphere $\Sigma_0$  with four marked points with cross-ratio $v$, 
joined in pairs by two very long tubes of proper length $\tau_2$ and $\sigma_2$. 
It turns out\footnote{Indeed, the local modular Laurent polynomials $A_{i,j}(S)$
arising in the limit $V\to 0$ 
are integrable around $S_1=0$, even though they are not smooth on this locus.}  that for the integrands of interest in this paper, this region does not contribute
additional divergent terms beyond those caused by the `minimal' and `maximal'
non-separating
degenerations, discussed above and below.

\item[iv)] The `maximal' or `complete'  non-separating
degeneration corresponds to setting $L_1,L_2$, $L_3\to\infty$ at the same rate, or equivalently $V\to 0$, keeping $\Re\Omega$ fixed. In this limit, $\Sigma_2$
degenerates into two spheres with three marked points on each, joined in pair by three propagators
of proper length $L_i$ so as to form a `sunset' graph (see diagram $(a)$ on Figure \ref{fig:twoloop}). 
This region is responsible for primitive two-loop divergences.
\end{itemize}
The regularization \eqref{defF2L} of the  fundamental domain amounts to imposing
an upper bound $t < L$ on the length for the long tube in the degeneration ii), and an 
upper bound $L_1<L_2<L_3<L$ on the proper lengths of the three parameters in 
degeneration iv). It does not regulate the separating degeneration i), but as already
mentioned, there are no divergences coming from this region in the cases of interest
in this paper.

\subsubsection*{The truncated modular integral}

To determine the $L$-dependence of the truncated modular integral \eqref{d8rfourreg}, it is
easiest to compute its derivative  with respect to $L$.
Since $L$ only affects the upper bound of the $t$ integral, the $L$-derivative is obtained
by evaluating the integrand at $t=L$,
\be
L\partial_L\, \cE^{(d,2)}_{(p,q)}(\varphi,L) = \frac{\pi}{4} \frac{2}{L^2}\,
 \int_{[-\frac12,\frac12]^3} \de u_1 \,\de u_2 \,\de \sigma_1\, 
 \int_{\cF_1\left(\frac{L}{1-u_2^2}\right)}  \de\mu_1(\tau)
\left[ \Gamma_{d,d,2}(\varphi)\,\cB^{(2)}_{(p,q)} \right]_{t=L}, 
\ee
The prefactor $2/L^2$ comes from the measure $\de\mu_2=4 \frac{\de t}{t^3} \de\mu_1(\rrho) \de u_1 \de u_2 \de\sigma_1$ and the $\IZ_2$ symmetry acting on $u_1,u_2,\sigma_1$, while
the restriction to the domain $\cF_1({L}/{(1-u_2^2)})$ comes from the constraint $\tau_2<\sigma_2=t+\tau_2 u_2^2 $
defining the fundamental domain $\cF_2$. Since $|u_2|<\frac12$, one has $\tau_2<4L/3$ 
therefore the integral over $\tau$ is finite. 

In order to disentangle the dependence on $L$ coming from the integrand and from the integration domain, we
 decompose\footnote{The domain $\cF_1(L_0)$ and $ \cS_1(L_0,L)$ correspond to the boundaries at $t=L$ of regions I and II, in the terminology of \cite{Pioline:2015nfa,Florakis:2016boz}.}
 $\cF_1({L}/{(1-u_2^2)})= \cF_1(L_0)\sqcup \cS_1(L_0,L)$,
 where $\cF_1(L_0)$ is the truncated fundamendal domain \eqref{defF1L}, while
 \be
 \cS_1(L_0,L)=\left\{ \tau\in \cH_1\ : \  L_0<\tau_2<\frac{L}{1-u_2^2}, \ -\frac12<\tau_1<\frac12\right\}\ .
 \ee
Here, $L_0$ is an arbitrary separating scale, chosen such that $1\ll L_0 \ll L$. 
For the domain $ \cS_1(L_0,L)$, we use the coordinates $V=1/\sqrt{t\tau_2}, 
S_2=\sqrt{t/\tau_2}, S_1=u_2$, so that the conditions defining the domain $ \cS_1(L_0,L)$
translate into 
\be
|S|>1\ ,\quad S_2<\sqrt{L/L_0}\ ,\quad V=\frac{S_2}{L}\ .
\ee
Using $\de\mu_2=8V^2\de V\de\mu(\ttau)\, \de \rho_1 \de v_1 \de \sigma_1$, with
$\de\mu(\ttau) = 2 \frac{\de \ttau_1\de\ttau_2}{\ttau_2^2}$, we obtain
\bea
L\partial_L\, \cE^{(d,2)}_{(p,q)}(\varphi,L) &=& \frac{\pi}{2L^2}\,
\int_{\cF_1(L_0)} \de\mu_1(\tau) \int_{[-\frac12,\frac12]^3} \de u_1 \,\de u_2 \,\de \sigma_1\,
\left[ \Gamma_{d,d,2}(\varphi)\,\cB^{(2)}_{(p,q)} \right]_{t=L} \\
&+& 2\pi \int_{\cF(\sqrt{L/L_0})} \de\mu(S) \int_{[-\frac12,\frac12]^3} \de \tau_1 \,\de v_1 \,\de \sigma_1\,
\left(\frac{S_2}{L}\right)^3
\left[ \Gamma_{d,d,2}(\varphi)\,\cB^{(2)}_{(p,q)}\right]_{V=S_2/L}
\nn
\ee
where $\cF(\sqrt{L/L_0})$ is the truncated fundamental domain for the action \eqref{GL2act} of $GL(2,\IZ)$
on the modulus $S$, namely
\be
\cF(\sqrt{L/L_0}) = \{ S \in \cH_1\ :  |S|>1, \ S_2 < \sqrt{L/L_0}\ ,\quad 0< S_1 < \frac12 \}
\ee  

For large $t=L$, we can approximate the  genus-two Narain partition function $\Gamma_{d,d,2}(\varphi)$ by  
$t^{d/2}\,\Gamma_{d,d,1}$ with exponential accuracy. The remaining integral over $u_1,u_2,\sigma_1$ picks up the zeroth Fourier-Jacobi coefficient of $\cB^{(2)}_{(p,q)}$. We shall
assume that for $t\to \infty$, this constant term is given, up to exponential corrections,
 by a finite Laurent series
\be
\label{BgrowthI}
\int_{[-\frac12,\frac12} \de u_1\, \de u_2 \, \de \sigma_1\,
\cB^{(2)}_{(p,q)}(\Omega) = \sum_i  t^{\sigma_i} \,\hat\cB_i(\rrho) + \cO(e^{-2\pi t})
\ee
where $\hat\cB_i(\rrho)$ are  modular functions of $\rrho$ with polynomial
growth as $\rrho_2\to \infty$. Indeed, in the cases of interest, $\cB^{(2)}_{(p,q)}$ itself has a 
finite Laurent expansion of the form \eqref{BgrowthI}, with $\sigma_i\in [-w,\dots, w]$
where $w=2p+3q-2$, whose coefficients 
$\hat\cB_i(\tau,v)$ are generalized modular graph functions of weight
$w-\sigma_i$~\cite{DHoker:2017pvk}. 

Similarly, for small $V=S_2/L$, we can approximate the Narain partition function $\Gamma_{d,d,2}(\varphi)$ by  
$V^{-d}$ with exponential accuracy. The remaining integral over $\tau_1,v_1,\sigma_1$ picks up the
zeroth Fourier coefficient of $\cB^{(2)}_{(p,q)}$. We shall assume that in the limit $V\to 0$, 
his constant term is given, up to exponentially small corrections,
 by a finite Laurent series,
\be
\label{BgrowthII}
\int_{[-\frac12,\frac12]^3} \de \tau_1 \de v_1 \de \sigma_1\,
\cB^{(2)}_{(p,q)}(\Omega) = \sum_j  V^{\alpha_j} \tilde\cB_j(\ttau) + \cO(e^{-1/\sqrt{V}})\ ,
\ee
 where $\tilde\cB_j(\ttau)$ are functions\footnote{Note that these functions need not be 
differentiable on the locus $\ttau_1=0$ and its images under $GL(2,\IZ)$, whenever the integrand
$\cB^{(2)}_{(p,q)}(\Omega)$
is singular in the separating degeneration. 
} on $\cH_1$, invariant under the action \eqref{GL2act}  of $GL(2,\IZ)$,
with polynomial growth as $\ttau_2\to\infty$. Indeed, in the cases of interest,
$\cB^{(2)}_{(p,q)}$ has a Laurent expansion of the form \eqref{BgrowthII} with 
$\alpha_i\in [-w, \dots, 2w]$, whose coefficients are modular local Laurent  
polynomials~\cite{DHoker:2017pvk}, a class of modular invariant but non smooth
functions on the Poincar\'e upper half plane
 (see Appendix \ref{sec_local} for a definition and properties of these functions). 

Under the assumptions \eqref{BgrowthI} and \eqref{BgrowthII},  we easily obtain
\bea
L\partial_L\, \cE^{(d,2)}_{(p,q)}(\varphi,L) &=&\frac{\pi}{2} \sum_i  L^{\sigma_i + \frac{d}{2}-2}\,
\int_{\cF_1(L_0)} \de\mu_1(\tau) \, \Gamma_{d,d,1}\,\hat\cB_i  \nn \\
&+& 2\pi \sum_j L^{d-\alpha_j-3}\, \int_{\cF(\sqrt{L/L_0})} \de\mu(S) \, 
S_2^{3+\alpha_j-d}\, \tilde\cB_j 
\eea
Assuming that the zero-mode $\int_0^1 \de\rrho_1\, \hat\cB_i(\rrho)$ 
grows like $\sum_j \hat c_{i,j}\, \tau_2^{\eta_{i,j}}$ as $\rrho_2\to\infty$, and that
the zero-mode $\int_0^{1/2} \de\ttau_1\, \tilde\cB_j(\ttau) $ grows as $\frac12\sum_i \tilde c_{j,i} \ttau_2^{\beta_{j,i}}$ as $\ttau_2\to\infty$, we can replace the integrals of 
$\cF_1(L_0)$ and over $\cF(\sqrt{L/L_0})$ by their renormalized values
\be
\label{intF1ren}
\int_{\cF_1(L_0)} \de \mu_1\, 
 \Gamma_{d,d,1}  \, \hat\cB_i 
= \int_{\cF_1} \de \mu_1\,
\Gamma_{d,d,1}\, \hat\cB_i
 +2 \sum_j \hat c_{i,j} \frac{L_0^{\eta_{i,j}+\frac{d}{2}-1}}{\eta_{i,j}+\frac{d}{2}-1} +\dots
\ee
\be
\label{intFren}
\int_{\cF(\sqrt{\frac{L}{L_0}}) }  \de \mu \,
S_2^{3-d-\alpha_j}  \tilde\cB_j  =
\int_{\cF } \de \mu\,
S_2^{3-d-\alpha_j}  \tilde\cB_j + \sum_i \tilde c_{j,i} 
\frac{\left({L}/{L_0}\right)^{\frac12(\alpha_j+\beta_{j,i}+2-d)}}{\alpha_j+\beta_{j,i}+2-d} 
+ \dots
\ee
where the dots indicate exponentially suppressed corrections as $L_0\gg 1$ and 
$L\gg L_0$, respectively.
As a result, we obtain
\bea
L\partial_L\, \cE^{(d,2)}_{(p,q)}(\varphi,L) &=&\frac{\pi}{2} \sum_i  L^{\sigma_i + \frac{d}{2}-2}\,
\int_{\cF_1} \de\mu_1 \, \Gamma_{d,d,1}\,\hat\cB_i  
+ 2\pi \sum_j L^{d-\alpha_j-3}\, \int_{\cF} \de\mu \, 
S_2^{3+\alpha_j-d}\, \tilde\cB_j \nn\\
&+& \pi \sum_{i,j} \hat c_{i,j} \frac{L^{\sigma_i + \frac{d}{2}-2} L_0^{\eta_{i,j}+\frac{d}{2}-1}}
{\eta_{i,j}+\frac{d}{2}-1}+
2\pi \sum_{j,i} \tilde c_{j,i} \frac{L^{d-\alpha_j-3}\, (L/L_0)^{\frac12(\alpha_j+\beta_{j,i}+2-d)}}
{\alpha_j+\beta_{j,i}+2-d} \nn \\
\eea
By matching the expansions at $\rrho_2\to\infty$ and $\ttau_2\to\infty$,
one easily shows that the coefficients $\hat c_{i,j}$ and $\tilde c_{j,i}$ are equal 
whenever $\eta_{i,j}=-\alpha_j-\sigma_i$ and 
 $\beta_{j,i}=\alpha_j+2\sigma_i$, so that the terms on the last line cancel.
 Integrating with respect to $L$, we find that in the limit $L\to \infty$,
 $\cE^{(d,2)}_{(p,q)}(\varphi,L)$ grows like
 \bea
e^{(d,2)}_{(p,q)}(\varphi,L) &=&\frac{\pi}{2} \sum_{\sigma_i<2-\frac{d}{2}} \frac{L^{\sigma_i + \frac{d}{2}-2}
}{\sigma_i + \frac{d}{2}-2}\,
\int_{\cF_1} \de\mu_1 \, \Gamma_{d,d,1}\,\hat\cB_i 
+ \frac{\pi}{2} \log L\, \sum_{\sigma_i= 2-\frac{d}{2}} \, 
 \int_{\cF_1} \de \mu_1\, \hat\cB_i \, \Gamma_{d,d,1}  \nn \\
&+& 2\pi \sum_{\alpha_j<d-3} \frac{L^{d-\alpha_j-3}}{d-\alpha_j-3}\, \int_{\cF} \de\mu\, 
S_2^{3+\alpha_j-d}\, \tilde\cB_j 
\nn \\
&+& 2\pi \log L\, \sum_{\alpha_j=d-3} \int_{\cF } \de \mu\, \tilde\cB_j
+ 2\pi (\log L)^2 \, \sum_{\alpha_j=d-3\atop \beta_{j,i}=-1} \tilde c_{j,i}  \ ,
 \eea
 up to an additive constant (and a constant multiple of $\log L$ when $(\log L)^2$ terms
 are present). 
 The `renormalized' integral is then defined by subtracting both divergent pieces
before taking the limit $L\to \infty$ \cite{Pioline:2015nfa,Florakis:2016boz},
\be
\label{d8rfourreg2}
\cE^{(d,2)}_{(p,q)}(\varphi)
=\lim_{L\to\infty} \left[ \cE^{(d,2)}_{(p,q)}(\varphi,L) -e^{(d,2)}_{(p,q)}(\varphi,L)   \right]\ .
\ee


\subsubsection*{Regulated genus-two effective interactions}
We now apply the previous prescription to the genus-two effective couplings \eqref{d8rfourreg}. In the limit
$t\to \infty$, the zero-mode of the integrands \eqref{Bpq} 
with respect to $u_1,u_2,\sigma_1$ behave as \cite{DHoker:2018mys} \footnote{\label{foozero}
To obtain these results, we start from Eq. (3.16) and (3.22) in \cite{DHoker:2018mys}, and use
the identities
\be
\int_{\Sigma_1} \kappa_1(v)\, g_a(v) = 0\ ,\quad 
\int_{\Sigma_1} \kappa_1(v)\, g_a(v) g_b(v)  = E_{a+b},  \quad
\int_{\Sigma_1} \kappa_1(v)\, D_a^{(1)} = 0\ ,\quad 
\ee
where $\kappa_1(v)=\frac{\I}{2\tau+2}\de v \wedge \de\bar v$ and $g_a(v), D_a^{(1)}(v)$
are the Kronecker-Eisenstein series defined in Eq. (3.25), (3.29) in loc. cit., along with the identity
$D_3=E_3+\zeta(3)$ from \cite[(3.11)]{D'Hoker:2015foa}.
}
\bea
\label{nonkz}
 \int_{[-\frac12,\frac12]^3} \de u_1\, \de u_2 \, \de \sigma_1\, \cB^{(2)}_{(0,1)} &=& \frac{2\pi t}{3} + \frac{5E_2}{\pi t} + \cO(e^{-2\pi t})\ .
\\
 \int_{[-\frac12,\frac12]^3} \de u_1\, \de u_2 \, \de \sigma_1\, \cB^{(2)}_{(2,0)}(\Omega) &=& 
\frac{8\pi^2}{45} t^2 + \frac{10}{3} E_2 
+ \frac{5E_3+3\zeta(3)}{2\pi t} +\frac{\hat\cB^{(2)}_{(2,0)}}{t^2}+ \cO(e^{-2\pi t})\ ,
\ee
where we have not attempted to compute the coefficient $\hat\cB^{(2)}_{(2,0)}$ of the $\cO(1/t^2)$
term.
As we shall see momentarily, the fact that the Laurent coefficients involve the same Eisenstein series $E_2, E_3$
that appear in the integrands \eqref{Jpq} of the genus-one amplitudes is not a coincidence. 
In the limit $V\to 0$, the zero-modes of $\cB^{(2)}_{(p,q)}$ with respect to $\Re(\Omega)$ instead behave as \cite{Pioline:2015qha,DHoker:2018mys} 
\bea
\label{tropkz}
 \int_{[-\frac12,\frac12]^3} \de^3\Omega_1\, \cB^{(2)}_{(0,1)} &=& 
 \frac{10\pi}{3V} A_{10} +  \frac{5\zeta(3)}{\pi^2} V^2  + \cO(e^{-1/\sqrt{V}})
 \\
 \label{tropkz2}
 \int_{[-\frac12,\frac12]^3} \de^3\Omega_1\,\cB^{(2)}_{(2,0)}&=&\frac{\pi^2}{V^{2}}\left[ -\frac{26}{189} A_{0,0} + \frac{20}{189} A_{0,2}+\frac{20}{99} A_{1,1} + \frac{64}{45} A_{2,0} \right]
\\&&
+\frac{V\, \zeta(3)}{\pi}\ \left[ \frac{9}{5}  A_{0,1} + \frac{8}{3} A_{1,0} \right]
+\frac{3\zeta(5) }{2\pi^3}V^3 A_{0,1} + \beta \frac{\zeta(3)^2}{\pi^4} V^4 + \cO(e^{-1/\sqrt{V}}) \nn
\eea
where $A_{i,j}(\ttau)$ are the local modular functions defined in Appendix \S\ref{sec_local},
and the coefficient $\beta$ in the last line of \eqref{tropkz2} was left unevaluated
in \cite{DHoker:2018mys}\footnote{In the first version of this work, Eq. \eqref{tropkz2} was off
by a factor of 2, due to a mistake in  \cite[(5.18)]{DHoker:2018mys} which has now been 
corrected. The coefficient $\beta$ in  \eqref{tropkz2}  is $\frac{\beta+11}{16}$ in \cite{DHoker:2018mys}.}. As explained in this reference, the leading term in these
expansions reproduces the integrand of the two-loop supergravity amplitude, in 
terms of the Schwinger parameters $L_1,L_2,L_3$ introduced in \eqref{defVttau}.
One of the main aims of this note is to elucidate the origin of the subleading coefficients,
proportional to odd zeta values. 
 
Using these asymptotics, we can easily extract the divergent terms in the truncated modular
integrals \eqref{d8rfourreg}\footnote{For $d=0$, the modular integral $ \cE_{(0,1)}^{(d,2)} (\varphi)$
was computed in \cite{D'Hoker:2014gfa} using the Laplace eigenvalue  property of $\varphi_{\rm KZ}$. For $d>0$, differential equations for the renormalized
integral $ \cE_{(0,1)}^{(d,2)} (\varphi)$ were established in \cite{Basu:2015dqa,Pioline:2015nfa}. It
would be very interesting to establish similar results for higher derivative interactions.},
 \bea
 \label{E10L2}
\cE_{(1,0)}^{(d,2)}(\varphi,L) &=&  
4\pi \frac{L^{d-3}}{d-3}\, F_{00}(d-3) 
+ \frac{L^{\frac{d-4}{2}}}{\frac{d-4}{2}} \cE_{(0,0)}^{(d,1)}(\varphi)
 + \cE_{(1,0)}^{(d,2)} (\varphi)
\\
\cE_{(0,1)}^{(d,2)}(\varphi,L) &= & 
\frac{20\pi^2}{3}  \frac{L^{d-2}}{d-2} \, F_{10}(d-2)
   +  \frac{10\zeta(3)}{\pi} \frac{L^{d-5}}{d-5}  \, F_{00}(d-5) 
   \nn\\ 
& +&
\frac{\pi}{3}  \frac{L^{\frac{d-2}{2}}}{\frac{d-2}{2}} \cE_{(0,0)}^{(d,1)}(\varphi)
+ \frac{5}{2\pi} \frac{L^{\frac{d-6}{2}}}{\frac{d-6}{2}} \cE_{(1,0)}^{(d,1)}(\varphi)
+ \cE_{(0,1)}^{(d,2)}(\varphi)
 \label{E01L2}\\
 \cE^{(d,2)}_{(2,0)}(\varphi,L) 
 &=& 
  2\pi^3  \frac{L^{d-1}}{d-1}
  \left[ -\tfrac{26}{189} F_{00}(d-1) + \tfrac{20}{189} F_{02}(d-1) +\tfrac{20}{99} F_{11}(d-1)  + \tfrac{64}{45} F_{20}(d-1)  \right]
\nn\\&
+ & 2\zeta(3)  \frac{L^{d-4}}{d-4}
\left[  \tfrac{9}{5} F_{01}(d-4) +\tfrac{8}{3}F_{10}(d-4) \right]
+ \frac{3\zeta(5)}{\pi^2} \frac{L^{d-6}}{d-6} 
F_{01}(d-6)
 \nn\\ &
+&  \frac{2\beta \zeta(3)^2}{\pi^2}  \frac{L^{d-7}}{d-7} F_{00}(d-7)   
 +  \frac{4\pi^2}{45} \frac{L^{\frac{d}{2}}}{\frac{d}{2}}  \cE_{(0,0)}^{(d,1)}(\varphi)
+  \frac{5}{3}   \frac{L^{\frac{d-4}{2}}}{\frac{d-4}{2}}   \cE_{(1,0)}^{(d,1)}(\varphi)
    \nn\\
&+&
   \frac{1}{4\pi}   \frac{L^{\frac{d-6}{2}}}{\frac{d-6}{2}} 
 \left[ 3\cE_{(0,1)}^{(d,1)} + 2\zeta(3)\, \cE_{(0,0)}^{(d,1)}(\varphi) \right] 
+ \frac{\pi}{2}  \frac{L^{\frac{d-8}{2}}}{\frac{d-8}{2}} \int_{\cF_1} \de\mu_1\, \hat\cB_{(2,0)}^{(2)}
 +  \cE_{(2,0)}^{(d,2)}(\varphi)
  \label{E20L2}
\eea
where $F_{ij}(s)$ denotes the integral of $A_{i,j}$  over the fundamental domain of $GL(2,\IZ)$,
\be
\label{Intttau}
 F_{ij}(s) = \int_{\cF} \de\mu(\ttau)\, \ttau_2^{-s} \, A_{i,j}(\ttau)\ .
\ee
As discussed in Appendix \ref{sec_local}, this  integral converges when $\Re(s)>i+j-1$, and extends to a meromorphic function
in the whole $s$-plane, with poles at integers in the range $[-3i-j-1,\dots, i+j-1]$. 
As usual, the factors of $L^{d-k}/(d-k)$ when $d=k$ must be treated by replacing $d$ by $d+2\epsilon$
and taking the limit $\epsilon\to 0$ after dropping polar terms, keeping in mind that additional poles can come from the integral $F_{ij}(d-k)$. Again, the fact that
the coefficients in these expansions involve the same odd zeta values and
(renormalized) modular integrals which occur at tree-level \eqref{Epqtree}  and genus-one  
\eqref{Epq1reg}, \eqref{Jpq}
is not a coincidence, 
and will be explained in Section \ref{sec_sugra}.

\subsection{Genus three}
Although the main emphasis of this note is on the genus one and two contributions, it will
be useful to recall the leading term in the genus three amplitude, computed using the
pure spinor formalism~\cite{Berkovits:2005df}. The genus three amplitude  contributes first
at order  $\nabla^6\cR^4$ in the derivative expansion, with a  coefficient  given 
by~\cite{Gomez:2013sla}
\be
\cE^{(d,3)}_{(0,1)}(\varphi,L)= \frac{5}{16} \int_{\cF_3(L)} \de\mu_3 \, \Gamma_{d,d,3}\ .
\ee 
Using the truncated fundamental domain $\cF_3(L)$ defined in \cite{Florakis:2016boz}, 
and specializing to the dimensions where logarithmic terms arise,  we get
(Eq. (4.72) in \cite{Florakis:2016boz}, noting that the measure $\de\mu_{h}$ used in that
reference differs from the one used here by a factor $1/2^{h(h+1)/2}$)
\bea
\cE^{(d,3)}_{(0,1)}(\varphi,L) &=&  \cE^{(d,3)}_{(0,1)}(\varphi)
+  \frac{5}{4} \log L\, \delta_{d,6} \, 
\int_{\cF_2} \de\mu_2 \, \Gamma_{6,6,2} \nn \\
&&
+  \frac{5\pi}{3} \log L\, \delta_{d,5} 
\int_{\cF_2} \de\mu_1 \, \Gamma_{5,5,1}
+ 5\zeta(3)\, \log L\,  \delta_{d,4}
\\
&=&  \cE^{(d,3)}_{(1,0)}(\varphi) + \frac{5}{2\pi}  \cE^{(6,2)}_{(1,0)}(\varphi) \, \log L\, \delta_{d,6} 
\nn\\&&+ \frac{5}{3} \cE^{(5,1)}_{(0,0)}(\varphi)\, \log L\, \delta_{d,5} + 5\zeta(3)\, \log L\,  \delta_{d,4}
 \label{E01L3}
\eea
The second term in the last line comes from a one-loop subdivergence in $D=4$, where the other two loops shrink to give a $\nabla^4\cR^4$ vertex. The third term comes from a two-loop subdivergence in $D=5$, where the 
remaining loop shrinks to give a $\cR^4$ vertex. The last term comes from the primitive three-loop divergence in $D=6$, and is proportional to  the volume of the fundamental domain of $PGL(3,\IZ)$, given by\footnote{As far as I know, $\zeta(3)$ here  is  completely unrelated to the
 coefficient of the tree-level $\cR^4$ interaction !} $\zeta(3)/24$. 
 
 Unfortunately, the genus-three contributions to higher derivative interactions $\cE_{(p,q)}^{(d,3)}$
 with $2p+3q>3$ are not known at present, due to difficulties in regulating the integral over the pure spinor ghosts \cite{Gomez:2013sla}. However, by requiring that coefficients of divergences recombine into U-duality invariant combinations, we shall see that the next-to-leading term
 must produce the following divergences in $D=6$ and $D=4$,
 \be
 \begin{split}
 \cE^{(d,3)}_{(2,0)}(\varphi,L) =&  \left[ \frac53  \cE^{(4,2)}_{(1,0)} \log L 
 + \frac76 \cE^{(4,2)}_{(1,0)} \, (\log L)^2  \right]  \delta_{d,4} \\
+& \left[   \frac{1}{4\pi} \left( 3  \cE_{(0,1)}^{(6,2)}
+ \frac12  (\cE_{(0,0)}^{(6,1)})^2 \right) \log L 
+  \frac{3}{2\pi^2} \cE_{(1,0)}^{(6,1)}\,(\log L )^2 \right] \delta_{d,6}  + \dots
 \label{E20L3}
 \end{split}
 \ee
In particular, the integrand  $\cB_{(2,0)}^{(3)}$ must produce the genus-two Kawazumi-Zhang invariant $\varphi_{KZ}$ in the limit where the genus-three curve degenerates into a genus-two curve with two marked points joined by a long tube. 

\section{Matching with supergravity divergences}
\label{sec_sugra}

The appearance of odd zeta values $\zeta(3), \zeta(5), \zeta(3)^2, \dots$ in
the asymptotics of the genus one and two integrands, as well as in the resulting
truncated modular integrals, strongly suggests that these terms should be related
to the tree-level $\cR^4, \nabla^4\cR^4, \nabla^6\cR^4$ interactions whose coefficients
 \eqref{Epqtree} involve the same transcendental numbers~\cite{DHoker:2018mys}. In
 this section, we argue that these terms are indeed related, by matching the 
logarithmic terms in the truncated modular integrals with ultraviolet divergences 
of loop amplitudes in  supergravity supplemented with higher derivative vertices.

\subsection{Qualitative analysis of UV divergences in supergravity}
Although supergravity amplitudes are most efficiently studied using dimensional
regularization, it is instructive to discuss their divergences using a hard UV cut-off $|k|<\Lambda$
on the Euclidean momenta of the particles running in the loop (assuming that such a cut-off
is consistent with supersymmetry). We are mostly interested in dimensions $D$
where logarithmic divergences occur, which corresponds to appearance of poles in 
dimensional regularization.

\subsection*{One-loop}

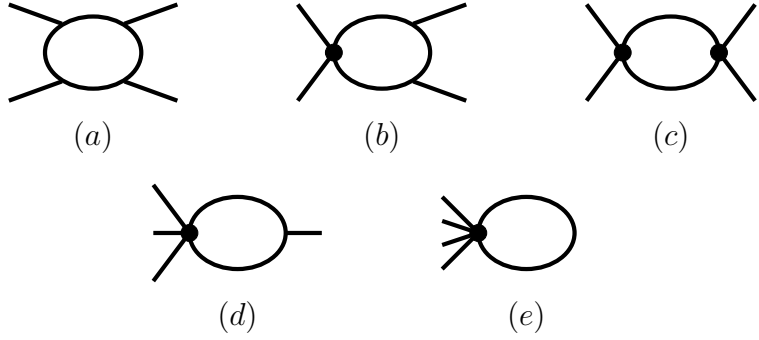
\begin{figure}[h]
\begin{center}
\begin{tikzpicture}[scale=.16]
\begin{scope}[shift={(-24,15)}]  
\draw [ultra thick] (0,0) ellipse (4 and 3);
\draw [ultra thick] (2.6,-2.4) -- (7,-4);
\draw [ultra thick] (2.6,2.4) -- (7,4);
\draw [ultra thick] (-2.6,-2.4) -- (-7,-4);
\draw [ultra thick] (-2.6,2.4) -- (-7,4);
\draw (0, -7.0) node{$(a)$};
\end{scope}
\begin{scope}[shift={(0,15)}]
\filldraw [black]  (-4,0) ellipse (0.7 and 0.7);
\begin{scope}[rotate=180]   
\end{scope}
\draw [ultra thick] (0,0) ellipse (4 and 3);
\draw [ultra thick] (2.6,-2.4) -- (7,-4);
\draw [ultra thick] (2.6,2.4) -- (7,4);
\draw [ultra thick] (-4,0) -- (-7,-4);
\draw [ultra thick] (-4,0) -- (-7,4);
\draw (0, -7.0) node{$(b)$};
\end{scope}
\begin{scope}[shift={(24,15)}]
\filldraw [black]  (4,0) ellipse (0.7 and 0.7);
\filldraw [black]  (-4,0) ellipse (0.7 and 0.7);
\draw [ultra thick] (0,0) ellipse (4 and 3);
\draw [ultra thick] (4,0) -- (7,-4);
\draw [ultra thick] (4,0) -- (7,4);
\draw [ultra thick] (-4,0) -- (-7,-4);
\draw [ultra thick] (-4,0) -- (-7,4);
\draw (0, -7.0) node{$(c)$};
\end{scope}
\begin{scope}[shift={(-12,0)}]
\filldraw [black]  (-4,0) ellipse (0.7 and 0.7);
\draw [ultra thick] (0,0) ellipse (4 and 3);
\draw [ultra thick] (-4,0) -- (-7,0);
\draw [ultra thick] (4,0) -- (7,0);
\draw [ultra thick] (-4,0) -- (-7,-4);
\draw [ultra thick] (-4,0) -- (-7,4);
\draw (0, -7.0) node{$(d)$};
\end{scope}
\begin{scope}[shift={(12,0)}]
\filldraw [black]  (-4,0) ellipse (0.7 and 0.7);
\draw [ultra thick] (0,0) ellipse (4 and 3);
\draw [ultra thick] (-4,0) -- (-7,3);
\draw [ultra thick] (-4,0) -- (-7,1);
\draw [ultra thick] (-4,0) -- (-7,-1);
\draw [ultra thick] (-4,0) -- (-7,-3);
\draw (0, -7.0) node{$(e)$};
\end{scope}
\end{tikzpicture}
\end{center}
\caption{$(a)$ Box diagram in supergravity; $(b)$ Triangle diagram with one $\nabla^{4p+6q} \cR^4$ quartic vertex; $(c)$  Bubble diagram with two quartic  vertices; $(d)$ Bubble diagram with one quintic and one cubic  vertex;  
$(e)$ Tadpole diagram with one sextic vertex.
\label{fig:oneloop}}
\end{figure}

By dimensional analysis,
the box diagram $(a)$ in Figure \ref{fig:oneloop} diverges  in dimension $D$ like
\be
\label{div1sug}
\int \de^D k \frac{(k^2)^4}{(k^2)^4} \sim \Lambda^{D-8} \, \cR^4 + \Lambda^{D-12} \nabla^4 \cR^4 
+ \Lambda^{D-14} \nabla^6 \cR^4 +\Lambda^{D-16} \nabla^8 \cR^4
+ \dots
\ee
where we used the fact that the leading term in the low energy expansion involves eight powers of momenta, which
combine with the eight polarization tensors into a $\cR^4$ vertex, while the next-to-leading
correction $\nabla^2 \cR^4$ vanishes on-shell.  Upon identifying $\Lambda=1/(\ell_s \sqrt{L})$, we observe that exactly the same powers of $L$ appear in the leading term in the
expansion of the genus-one truncated modular integrals \eqref{E00L1}--\eqref{E11L1}.
When $D=8$, the coefficient of $\cR^4$ becomes proportional to $\log \Lambda$,
in agreement with  the $\log L$ term in \eqref{E00L1} for $d=2$. The
higher-derivative terms in \eqref{div1sug} give logarithmic divergences in $D\geq 12$,
but those are unphysical since they occur above the critical dimension $D=10$. 

However, in the presence of higher-derivative tree-level vertices in the action, of the form 
$\int \de^D x\, \sqrt{-g}\, \cE_{(p,q)}^{(d,0)} \nabla^{4p+6q}(\varphi,L) \cR^4$, the one-loop amplitude also includes a triangle 
graph $(b)$ with one quartic vertex, and a bubble graph $(c)$ with two quartic vertices. For
$(p,q)=(0,0)$, the triangle graph $(b)$ leads to divergences of the form\footnote{This can be viewed as
 the  form factor of the supersymmetric completion of the operator $\cR^4$ 
  at one loop and zero momentum, as emphasized in \cite{Bossard:2015oxa}.}
\be
\label{div1r4}
\cE_{(0,0)}^{(d,0)}
\int\!\! \de^D k \frac{(k^2)^2 k^8}{(k^2)^3} \sim \cE_{(0,0)}^{(d,0)} \left[ 
\Lambda^{D-2} \, \cR^4 + \Lambda^{D-6} \nabla^4 \cR^4 
+ \Lambda^{D-8} \nabla^6 \cR^4 + \Lambda^{D-10} \nabla^8 \cR^4 + \dots \right]
\ee
The term proportional to $\Lambda^{D-6}$, $\Lambda^{D-8}$, $\Lambda^{D-10}$
lead to a logarithmic divergence in $D=6,8,10$,  corresponding
to the second term in \eqref{E10L1},\eqref{E01L1} and \eqref{E20L1}.
It is important to stress that in \eqref{div1r4}, we used the on-shell value of the
quartic vertex, whereas two legs are a priori off-shell. This does not affect the imaginary part
of the scattering amplitude in the forward limit, and therefore should preserve the structure
of the UV divergences, as we further elaborate in \S\ref{sec_discuss}.

Similarly, the triangle graph $(b)$ 
with one $\cE_{(1,0)}^{(d,0)}(\varphi,L)  \nabla^4\cR^4$ vertex produces
\be
\label{div1d4r4}
\cE_{(1,0)}^{(d,0)}
\int\!\! \de^D k \frac{(k^2)^2 k^{12}}{(k^2)^3} \sim \cE_{(1,0)}^{(d,0)} \left[ 
\Lambda^{D+2} \, \cR^4 + \Lambda^{D-2} \nabla^4 \cR^4 
+ \Lambda^{D-4} \nabla^6 \cR^4 + \Lambda^{D-6} \nabla^8 \cR^4 + \dots \right]
\ee
The terms proportional to $\Lambda^{D-4}$, $\Lambda^{D-6}$,  $\Lambda^{D-8}\dots$ lead to logarithmic divergences in $D=4, 6, 8,\dots$, correspond to the third term in \eqref{E01L1}, \eqref{E20L1}, and \eqref{E11L1}. The term proportional to $\Lambda^{D-2}$  indicates a divergence proportional to $\nabla^4\cR^4$ in $D=2$, but its physical significance is unclear due to the usual difficulties in defining graviton scattering in $D=2$.

In turn, the triangle graph $(b)$ with one $\nabla^6\cR^4$ vertex, along with the bubble graph (c)
with two $\cR^4$ vertices, produce divergences of the form 
 \be
 \label{div1d6r4}
\left[ \cE_{(0,1)}^{(d,0)} + (\cE_{(0,0)}^{(d,0)})^2 \right]
\left[ 
\Lambda^{D+4} \, \cR^4 + \Lambda^{D} \nabla^4 \cR^4 
+ \Lambda^{D-2} \nabla^6 \cR^4 + \Lambda^{D-4} \nabla^8 \cR^4 + \Lambda^{D-6} \nabla^{10} \cR^4 + \dots \right]
\ee
where the relative coefficient between $\cE_{(0,1)}^{(d,0)}$ and $(\cE_{(0,0)}^{(d,0)})^2$ is not
known.
The term proportional to $\Lambda^{D-4}$ correspond to the fourth term in \eqref{E20L1}, while the term proportional to $\Lambda^{D-6}$ correspond to the fourth term in \eqref{E11L1}. 

In addition, the tree-level action in string theory contains higher order gravitational vertices of the form
$\nabla^{2n}\cR^5$ with $n\geq 1$, $\nabla^{2n} \cR^6$ with $n\geq 0$, etc,  \cite{Stieberger:2009rr,Schlotterer:2012ny}, which can be 
used to construct the  bubble graph $(d)$ with one quintic vertex, or the  tadpole graph $(e)$ with one cubic vertex. The latter has no dependence on the external momenta and can be safely ignored.
As for the contributions of type $(d)$, at tree level they are related by non-linear 
supersymmetry to $\nabla^{4p+6q} \cR^4$ couplings with $4p+6q=2n-2$, in which case they produce the same
type of divergences as \eqref{div1d4r4}, \eqref{div1d6r4}, etc. Therefore there are no additional
contributions at first order in the genus expansion.

\subsection*{Two-loop}
\begin{figure}[h]
\begin{center}
\begin{tikzpicture}[scale=.16]
\begin{scope}[shift={(-15,0)}]
\draw [ultra thick] (-2.6,-2.4) -- (-7,-4);
\draw [ultra thick] (-2.6,2.4) -- (-7,4);
\draw [ultra thick] (2.6,-2.4) -- (7,-4);
\draw [ultra thick] (2.6,2.4) -- (7,4);
\draw [ultra thick] (0,0) ellipse (4 and 3);
\draw [ultra thick] (0,3) -- (0,-3);
\draw (0, -7.0) node{$(a)$};
\end{scope}
\begin{scope}[shift={(15,0)}]
\filldraw [black]  (-4,0) ellipse (0.7 and 0.7);
\begin{scope}[rotate=180]   
\end{scope}
\draw [ultra thick] (0,0) ellipse (4 and 3);
\draw [ultra thick] (2.6,-2.4) -- (7,-4);
\draw [ultra thick] (2.6,2.4) -- (7,4);
\draw [ultra thick] (-4,0) -- (-7,-4);
\draw [ultra thick] (-4,0) -- (-7,4);
\draw [ultra thick] (0,3) -- (0,-3);
\draw (0, -7.0) node{$(b)$};
\end{scope}
\begin{scope}[shift={(45,0)}]
\filldraw [black]  (0,0) ellipse (0.7 and 0.7);
\begin{scope}[rotate=180]   
\draw [ultra thick] (0,0) .. 
controls (-8,9) and (-8,-9) .. 
(0,0);
\end{scope}
\draw [ultra thick] (0,0) .. 
controls (-8,9) and (-8,-9) .. 
(0,0);
\draw [ultra thick] (-5,-2.4) -- (-7,-4);
\draw [ultra thick] (-5,2.4) -- (-7,4);
\draw [ultra thick] (5,-2.4) -- (7,-4);
\draw [ultra thick] (5,2.4) -- (7,4);
\draw (0, -7.0) node{$(c)$};
\end{scope}
\begin{scope}[shift={(-15,-20)}]
\filldraw [black]  (4,0) ellipse (0.7 and 0.7);
\filldraw [black]  (-4,0) ellipse (0.7 and 0.7);
\draw [ultra thick] (0,0) ellipse (4 and 3);
\draw [ultra thick] (0,3) -- (0,-3);
\draw [ultra thick] (4,0) -- (7,-4);
\draw [ultra thick] (4,0) -- (7,4);
\draw [ultra thick] (-4,0) -- (-7,-4);
\draw [ultra thick] (-4,0) -- (-7,4);
\draw (0, -7.0) node{$(d)$};
\end{scope}
\begin{scope}[shift={(15,-20)}]
\filldraw [black]  (-3.5,1.0) ellipse (0.7 and 0.7);
\filldraw [black]  (3.5,1.0) ellipse (0.7 and 0.7);
\draw [ultra thick] (0,0) ellipse (4 and 3);
\draw [ultra thick] (3.5,-1.5) -- (7,-4);
\draw [ultra thick] (3.5,1.0) -- (7,4);
\draw [ultra thick] (-3.5,-1.5) -- (-7,-4);
\draw [ultra thick] (-4,1.0) -- (-7,4);
\draw [ultra thick] (-3.5,1.0) -- (3.5,1.0);
\draw (0, -7.0) node{$(e)$};
\end{scope}
\begin{scope}[shift={(45,-20)}]
\filldraw [black]  (0,0) ellipse (0.7 and 0.7);
\filldraw [black]  (6,0) ellipse (0.7 and 0.7);
\begin{scope}[rotate=180]   
\draw [ultra thick] (0,0) .. 
controls (-8,9) and (-8,-9) .. 
(0,0);
\end{scope}
\draw [ultra thick] (0,0) .. 
controls (-8,9) and (-8,-9) .. 
(0,0);
\draw [ultra thick] (-5,-2.4) -- (-7,-4);
\draw [ultra thick] (-5,2.4) -- (-7,4);
\draw [ultra thick] (6,0) -- (9,-4);
\draw [ultra thick] (6,0) -- (9,4);
\draw (0, -7.0) node{$(f)$};
\end{scope}
\begin{scope}[shift={(-15,-40)}]
\filldraw [black]  (-4,0) ellipse (0.7 and 0.7);
\begin{scope}[rotate=180]   
\end{scope}
\draw [ultra thick] (0,0) ellipse (4 and 3);
\draw [ultra thick] (3.5,-1.5) -- (7,-4);
\draw [ultra thick] (3.5,1.5) -- (7,4);
\draw [ultra thick] (-4,0) -- (-7,-4);
\draw [ultra thick] (-4,0) -- (-7,4);
\draw [ultra thick] (-4,0) -- (4,0);
\draw (0, -7.0) node{$(g)$};
\end{scope}
\begin{scope}[shift={(15,-40)}]
\filldraw [black]  (-4,0) ellipse (0.7 and 0.7);
\filldraw [black]  (3.5,1.0) ellipse (0.7 and 0.7);
\begin{scope}[rotate=180]   
\end{scope}
\draw [ultra thick] (0,0) ellipse (4 and 3);
\draw [ultra thick] (3.5,-1.5) -- (7,-4);
\draw [ultra thick] (3.5,1.0) -- (7,4);
\draw [ultra thick] (-4,0) -- (-7,-4);
\draw [ultra thick] (-4,0) -- (-7,4);
\draw [ultra thick] (-4,0) -- (3.5,1.0);
\draw (0, -7.0) node{$(h)$};
\end{scope}
\begin{scope}[shift={(45,-40)}]
\filldraw [black]  (-4,0) ellipse (0.7 and 0.7);
\filldraw [black]  (4,0) ellipse (0.7 and 0.7);
\begin{scope}[rotate=180]   
\end{scope}
\draw [ultra thick] (0,0) ellipse (4 and 3);
\draw [ultra thick] (4,0) -- (7,-4);
\draw [ultra thick] (4,0) -- (7,4);
\draw [ultra thick] (-4,0) -- (-7,-4);
\draw [ultra thick] (-4,0) -- (-7,4);
\draw [ultra thick] (-4,0) -- (4,0);
\draw (0, -7.0) node{$(j)$};
\end{scope}
\end{tikzpicture}
\end{center}
\caption{$(a)$ Two-loop `sunset' diagram in supergravity (drawn sideways); 
$(b,c)$ Two-loop diagrams with one quartic vertex; 
 $(d,e,f)$  Two-loop diagrams with two quartic vertices;
$(g,h,j)$ Two-loop diagrams with quintic vertices; variants of these
diagrams with non-planar topologies are also allowed.}
\label{fig:twoloop}
\end{figure}
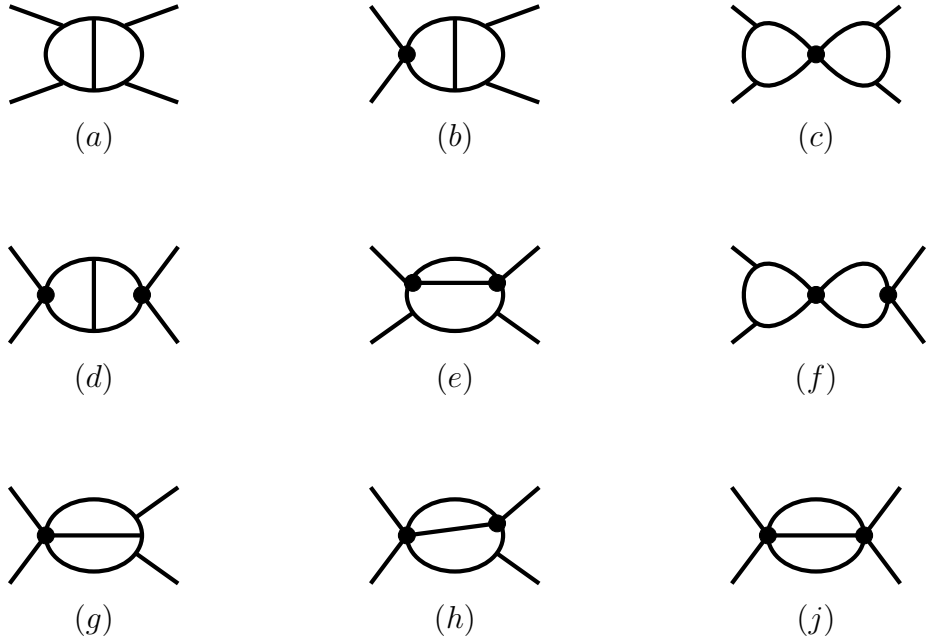

We now turn to the two-loop supergravity amplitude. By dimensional analysis, the two-loop sunset graph $(a)$ in Figure \ref{fig:twoloop} diverge like
\be
\int \de^{2D} k \frac{(k^2)^6}{(k^2)^7} \sim \Lambda^{2D-14} \, \nabla^4\cR^4 + \Lambda^{2D-16} \nabla^6 \cR^4 
+ \Lambda^{2D-18} \nabla^8 \cR^4
+\dots
\label{div2sug}
\ee
where we took into account the fact that the leading term in the low energy expansion starts at 
order $\nabla^4\cR^4$ \cite{Bern:1998ug}. After substituting $\Lambda=1/\sqrt{\alpha' L}$,
we see that the same powers of $L$ appear in the first term in the expansions 
\eqref{E10L2}--\eqref{E20L2}.
These terms generate logarithmic divergences in $D=7,8,9,\dots$, corresponding to 
the terms proportional to $L^{d-3}$ in \eqref{E10L2}, $L^{d-2}$ in \eqref{E01L2} 
and $L^{d-1}$ in \eqref{E20L2}, respectively. In addition to the primitive divergence \eqref{div2sug},
there are also one-loop subdivergences, obtained by inserting one of the one-loop supergravity divergences in \eqref{div1sug} at the quartic vertex of a triangle diagram $(b)$ in 
Figure \ref{fig:oneloop}. The resulting divergences are similar to \eqref{div1r4}--\eqref{div1d6r4},
upon replacing the tree-level coefficient $\cE_{(1,0)}^{(d,0)}$ is replaced by the divergent part of the
one-loop coefficient $\cE_{(1,0)}^{(d,1)}(L)$. In dimension $D=8$, this leads to a $(\log \Lambda)^2$
divergence, or a $1/\epsilon^2$ pole in dimensional regularization, in agreement with the supergravity
computation in 
\cite{Bern:1998ug} (see \eqref{div28} below).

\medskip

In addition, the diagrams $(b,c)$ in Figure \ref{fig:twoloop}, where two of the supergravity cubic vertices are replaced by  a tree-level $\cR^4$ vertex give divergences of the form
\be
\label{div2r4}
\cE_{(0,0)}^{(d,0)} \int \de^{2D} k \frac{(k^2)^4 k^8}{(k^2)^6} \sim \cE_{(0,0)}^{(d,0)} \left[ \Lambda^{2D-8} \, \nabla^4\cR^4 + \Lambda^{2D-10} \nabla^6 \cR^4 
+ \Lambda^{2D-12} \nabla^8 \cR^4
+\dots \right]
\ee
The second and third terms generate logarithmic singularities in $D=5$ and $D=6$, corresponding to   the terms proportional to $\zeta(3)\Lambda^{d-5}$ in \eqref{E01L2}
and to $\zeta(3)\Lambda^{d-4}$  in \eqref{E20L2}. The first term in \eqref{div2r4} 
would generate a logarithmic
divergence proportional to $\nabla^4\cR^4$ in $D=4$, but no such divergence appears in \eqref{E10L2}, so such a term is presumably ruled out by supersymmetry \footnote{Indeed,
the coefficients $\cE_{(0,0)}^{(d)}$ and $\cE_{(1,0)}^{(d)}$ have different eigenvalues under the
Laplace operator on $\cM_D$, so cannot mix. I am grateful to G. Bossard for pointing this out.}.  Similarly, if two of the supergravity cubic vertices are replaced by  a tree-level 
$\nabla^4\cR^4$ vertex, we get 
\be
\label{div2d4r4}
\cE_{(1,0)}^{(d,0)} \int \de^{2D} k \frac{(k^2)^4 k^{12}}{(k^2)^6} \sim \cE_{(1,0)}^{(d,0)} \left[ \Lambda^{2D-4} \, \nabla^4\cR^4 + \Lambda^{2D-6} \nabla^6 \cR^4 
+ \Lambda^{2D-8} \nabla^8 \cR^4
 +\dots \right]
\ee
The term proportional to $ \Lambda^{2D-8}$ generates a logarithmic divergence in  $D=4$, corresponding to the   term proportional to $\zeta(5) L^{d-6}$ in \eqref{E20L2}, while the term 
proportional to  $\Lambda^{2D-6}$ suggests a logarithmic divergence proportional
to $\nabla^6\cR^4$ in $D=3$, which is however not present in \eqref{E01L2}. Again,
such a term is presumably ruled out by supersymmetry, since the coefficients $\cE_{(1,0)}^{(d)}$ and $\cE_{(0,1)}^{(d)}$ for $d=7$ have different eigenvalues under the
Laplace operator on $\cM_D$.

Finally, replacing two of the supergravity vertices by a $\nabla^6\cR^4$ vertex
(still corresponding to the diagrams $(b,c)$ in Figure \ref{fig:twoloop}), or four supergravity vertices by two $\cR^4$ vertices (corresponding to the diagrams $(d,e,f)$ in Figure \ref{fig:twoloop}), we get 
in either case 
 \be
 \label{div2d6r4}
\left[ \cE_{(0,1)}^{(d,0)} + (\cE_{(0,0)}^{(d,0)})^2 \right]
\left[ \Lambda^{2D-2} \, \nabla^4\cR^4 + \Lambda^{2D-4} \nabla^6 \cR^4 
+ \Lambda^{2D-6} \nabla^8 \cR^4 
+\dots 
\right]
\ee
The term proportional to $\Lambda^{2D-6}$ generates a logarithmic singularity in $D=3$
corresponding to the term proportional to $\zeta(3)^2 L^{d-7}$ in \eqref{E20L2}. 

Still, this does not exhaust the set of two-loop diagrams which occur in the low energy limit of string theory. First, in addition 
to the tree-level  $\nabla^{4p+6q} \cR^4$ quartic vertices, the tree-level quintic vertices of the 
form $\nabla^{2n}\cR^5$ with $n\geq 1$ can be used to construct the two-loop diagrams $(g,h,j)$
on Figure \ref{fig:twoloop}. Power counting shows that the diagrams $(h,j)$  superficially diverge as  
$\Lambda^{2D+14}$ and $\Lambda^{2D+18}$ or faster, so do not lead to logarithmic 
divergences in dimension $D\geq 3$ up
to order $\nabla^{10}\cR^4$ included. In contrast, the diagram $(g)$ diverges as 
$\Lambda^{2D+2n-6} \nabla^4\cR^4$;  this is the same scaling as \eqref{div2d4r4} or 
\eqref{div2d6r4} for $n=1$ or $n=2$, in agreement with the fact that the $\nabla^{2n}\cR^5$
tree-level coupling is related by non-linear supersymmetry to the $\nabla^{2n-2} \cR^4$ couplings. 
Thus there are no additional divergences from diagrams $(g,h,j)$.

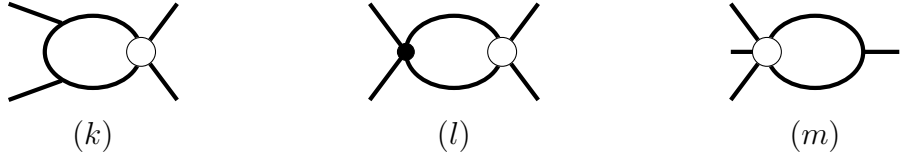
\begin{figure}[h]
\begin{center}
\begin{tikzpicture}[scale=.16]
\begin{scope}[shift={(-30,0)}]
\begin{scope}[rotate=180]   
\end{scope}
\draw [ultra thick] (0,0) ellipse (4 and 3);
\draw [ultra thick] (-2.6,-2.4) -- (-7,-4);
\draw [ultra thick] (-2.6,2.4) -- (-7,4);
\draw [ultra thick] (4,0) -- (7,-4);
\draw [ultra thick] (4,0) -- (7,4);
\filldraw [white]  (4,0) ellipse (1.2 and 1.2);
\draw [black]  (4,0) ellipse (1.2 and 1.2);
\draw (0, -7.0) node{$(k)$};
\end{scope}
\begin{scope}[shift={(0,0)}]
\draw [ultra thick] (4,0) -- (7,-4);
\draw [ultra thick] (4,0) -- (7,4);
\draw [ultra thick] (-4,0) -- (-7,-4);
\draw [ultra thick] (-4,0) -- (-7,4);
\draw [ultra thick] (0,0) ellipse (4 and 3);
\filldraw [white]  (4,0) ellipse (1.2 and 1.2);
\draw [black]  (4,0) ellipse (1.2 and 1.2);
\filldraw [black]  (-4,0) ellipse (0.7 and 0.7);
\draw (0, -7.0) node{$(l)$};
\end{scope}
\begin{scope}[shift={(30,0)}]
\draw [ultra thick] (0,0) ellipse (4 and 3);
\draw [ultra thick] (-4,0) -- (-7,0);
\draw [ultra thick] (4,0) -- (7,0);
\draw [ultra thick] (-4,0) -- (-7,-4);
\draw [ultra thick] (-4,0) -- (-7,4);
\filldraw [white]  (-4,0) ellipse (1.2 and 1.2);
\draw [black]  (-4,0) ellipse (1.2 and 01.2);
\draw (0, -7.0) node{$(m)$};
\end{scope}
\end{tikzpicture}
\end{center}
\caption{One-loop diagrams with genus-one quartic or quintic vertex (denoted by a white circle)
\label{fig:oneloop1}}
\end{figure}

More importantly however, there are also triangle and bubble 
diagrams with an insertion of genus-one  vertices $\cE_{(p,q)}^{(d,1)} \nabla^{4p+6q}\cR^4$
(see Figure \ref{fig:oneloop1}). 
The diagrams $(k,l)$ can be analyzed in the same way as the one-loop diagrams in the previous
subsection, upon replacing $\cE_{(p,q)}^{(d,0)}$ by  $\cE_{(p,q)}^{(d,1)}$ in 
\eqref{div1r4}--\eqref{div1d6r4}. It is easy to check that they reproduce the one-loop divergences of the form $\Lambda^{(d-k)/2} 
\cE_{(p,q)}^{(d,1)}/\frac{d-k}{2}$ in \eqref{E10L2}--\eqref{E20L2}. In fact, combining
 this supergravity description with U-duality, we predict that 
these one-loop subdivergences must combine
with the   divergences in  \eqref{E10L1}--\eqref{E20L1} of the genus one couplings
so as to form a multiple of the U-duality invariant  coefficient $\cE_{(p,q)}^{(d)} \propto g_D^{-2} 
\cE_{(p,q)}^{(d,0)} + \cE_{(p,q)}^{(d,1)} + \dots$. Expanding this at the next order in $g_D$
also predicts the form of the one-loop and two-loop subdivergences in the genus-three 
amplitude, as we discuss below. We verify some of these predictions in subsection \ref{sec_combi}.

As for the diagram $(m)$ in Figure \ref{fig:oneloop1}, it leads to the same type of divergences as
the diagram $(d)$ in Figure \ref{fig:oneloop}, with the important difference that at one-loop,
some of the quintic vertices of the form $\nabla^{2n}\cR^5$ with  $n\geq 3$ are no longer 
related by non-linear supersymmetry to $\nabla^{4p+6q}\cR^4$ couplings  \cite{Green:2013bza}.
For $n=3$, we find a superficial divergence of order $\Lambda^{D+14}$, which may lead to a
logarithmic divergence at order $\nabla^{10}\cR^4$ in $D=4$, $\nabla^{12}\cR^4$ in $D=6$, etc,
with a new set of U-duality invariant coefficients. Since we mostly restrict to order $\nabla^{8}\cR^4$
in this note, we can ignore such contributions.

\medskip

\subsection*{Three-loop}

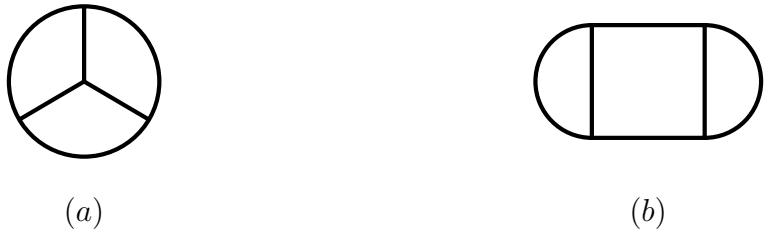
\begin{figure}[h]
\begin{center}
\begin{tikzpicture}[scale=.25]
\begin{scope}[shift={(-15,0)}]
\draw [ultra thick] (0,0) ellipse (4 and 4);
\draw [ultra thick] (0,0) -- (0,4);
\draw [ultra thick] (0,0) -- (3.4,-2);
\draw [ultra thick] (0,0) -- (-3.4,-2);
\draw (0, -7.0) node{$(a)$};
\end{scope}
\begin{scope}[shift={(15,0)}]
\draw [ultra thick] (-3,-3) rectangle ++(6,6);
\draw [ultra thick] (3,-3) arc (-90:90:3cm);
\draw [ultra thick] (-3,3) arc (90:270:3cm);
\draw (0, -7.0) node{$(b)$};
\end{scope}
\end{tikzpicture}
\end{center}
\caption{$(a)$ Skeleton of the  tetrahedron diagram; $(b)$ Skeleton of the three-loop
ladder diagram; in both cases, the four external gravitons can attach to any of the edges.
\label{fig:threeloop}}
\end{figure}

Finally, let us briefly consider three-loop diagrams. In supergravity,  they occur into two topologies, the tetrahedron (or `Mercedes') and ladder diagrams (see Figure \ref{fig:threeloop}). 
Under naive power counting, both lead to the same structure of primitive divergences,
\be
\label{div3sug}
\int \de^{3D} k \frac{(k^2)^8}{(k^2)^{10}} 
\sim \Lambda^{3D-18} \, \nabla^6\cR^4 + \Lambda^{3D-20} \nabla^8 \cR^4 
+ \Lambda^{3D-22} \nabla^{10}\cR^4 +\dots
\ee
where we took into account the fact that the leading term in the derivative expansion starts
at order $\nabla^6\cR^4$ (and in fact  comes only from the tetrahedron diagram) \cite{Bern:2008pv}. 
The first term in \eqref{div3sug} produces a logarithmic divergence in $D=6$, in agreement
with the three-loop divergence computed in \cite{Bern:2008pv} using dimensional regularisation.
In addition, there are also one-loop subdivergences, obtained by inserting one of the two-loop divergent terms
in \eqref{div2sug} at the quartic vertex of a triangle diagram $(b)$ in 
Figure \ref{fig:oneloop}; these contributions take the same form as \eqref{div1r4}--\eqref{div1d6r4},
upon replacing the tree-level  coefficient $\cE_{(1,0)}^{(d,0)}$ by the divergent part of the
one-loop term $\cE_{(1,0)}^{(d,1)}(L)$. Finally, there are also two-loop subdivergences, 
obtained by inserting one of the one-loop divergences
in \eqref{div1sug} at the quartic vertices of the diagrams $(b,c)$ in 
Figure \ref{fig:twoloop}; they take the same form as \eqref{div2r4}--\eqref{div2d6r4},
upon replacing $\cE_{(1,0)}^{(d,0)}$ by the divergent part of the
two-loop term $\cE_{(1,0)}^{(d,2)}(L)$. For $D=8, 10$, this produces $(\log\Lambda)^2$ and
$(\log\Lambda)^3$ terms, albeit at high order in the derivative expansion ($\nabla^{12}\cR^4$ and $\nabla^{18}\cR^4$, respectively).

Inserting higher derivative vertices, we get additional log divergences in $D=4$
and $D=2$ (and in fractional dimensions which we ignore):
\be
\cE_{(0,0)}^{(d)}\, \Lambda^{3D-12}\, \nabla^6\cR^4 + \cE_{(1,0)}^{(d)}\, \Lambda^{3D-12}\, \nabla^{10}\cR^4
+ \left[  \cE_{(0,1)}^{(d)}+ (\cE_{(0,0)}^{(d)})^2 \right]\, \Lambda^{3D-6}\, \nabla^6\cR^4 + \dots
\ee
Again, the logarithmic divergence proportional to $\nabla^6\cR^4$ in $D=4$ does not match any pole in \eqref{div6},
so its coefficient presumably vanishes.

\subsection{Combining genus-one and genus-two scale-dependent terms
\label{sec_combi}}
As explained above, dimensional analysis in supergravity supplemented with higher-derivative
stringy vertices gives a simple prediction for the dimensions $d$ where the truncated modular integrals
can have logarithmic divergences, and for the coefficients of these logarithmic divergences. In
particular, it requires that the coefficients of the genus-one divergence and of the one-loop
subdivergence of the genus-two divergence must recombine into the leading
terms in the U-duality invariant coupling $\cE_{(p,q)}^{(d)}$. In this subsection, we consider
each dimension in turn, and check that these predictions are indeed verified.

\begin{itemize}
\item In $D=10$, the logarithms
coming from the  divergence \eqref{E20L1} of the genus-one amplitude and the one-loop
subdivergence \eqref{E20L2} of the genus-two amplitude  combine into
\be
\label{div0}
\cE_{(2,0)}^{(0)}(\varphi,L) \sim 
g_s^{1/2} \left[ \frac{8 \pi^2\zeta(3)}{45} g_s^{-2} + \frac{4\pi^2}{45}   \cE_{(0,0)}^{(0,1)} +\dots \right] 
\log L
=  \frac{4\pi^2}{45}   \cE_{(0,0)}^{(0)}\log L
\ee
Here, the  overall power of
$g_D$ comes from rescaling from string to Einstein frame, and the dots indicate divergent contributions
from the instanton series, which are predicted by U-duality. As noted below \eqref{div1r4},
The contribution $\cE_{(0,0)}^{(0)}\log L$  can be understood as the divergence of a triangle diagram in supergravity with one
 $\cR^4$ quartic vertex involving both tree-level and one-loop contribution. While it would be desirable to recover 
 the coefficient $\frac{4\pi^2}{45}$ by an independent computation of the divergence of the
one-loop form factor in supergravity, this coefficient follows here  from a straightforward
analysis of 
 the subdivergence of the genus-two string amplitude.  At order $\nabla^{10}\cR^4$, 
there is also a two-loop supergravity divergence in $D=10$,
as discussed in Appendix~\ref{sec_d10r4}.

\item In $D=9$, there is a single divergent term up to order $\nabla^8\cR^4$, from 
\eqref{E20L2}.  Its coefficient is easily computed using \eqref{Fijzero},
\be
\label{div1}
\cE_{(2,0)}^{(1)}(\varphi,L) \sim -\frac{52 \pi^4}{567} \log  L 
\ee
According to the discussion below \eqref{div2sug}, this should come from the two-loop divergence \eqref{div29} in supergravity. In order to check the numerical factor, we quote the result of the two-loop supergravity divergence in $D=9$ \cite[(4.3)]{Bern:1998ug}
\be
\label{div29}
\cI_2^{D=9-2\epsilon} \sim -  \left(\frac{\kappa}{2}\right)^6\,  \frac{13\pi}{9072}\, \frac{1}{4\epsilon(4\pi)^9} \,(s^2+t^2+u^2)^2 \, stu M_4^{\rm tree}
\ee
To translate to our conventions, recall that $\sigma_n=(s^n+t^n+u^n)\ell_D^{2n}/4^n$,
while the Newton constant $\kappa^2$ is related to the Planck length
$\ell_D$ in dimension $D$ by\footnote{This differs by a factor 4 from 
the identification stated in \cite[\S B]{Green:2010sp}, which is necessary to
match the string divergences with results in the supergravity literature for all dimensions
and loop orders relevant in this paper -- with one possible exception, mentioned below \eqref{div2sugmul}. The matching of primitive divergences is in fact guaranteed by
the fact that the tropical
limit of the string integrand reproduces the supergravity integrand in Schwinger-type 
parametrization \cite{Tourkine:2013rda}. 
}
 $\kappa^2=2(2\pi)^{D-3} \ell_D^{D-2}$.
Identifying $stu M_4^{\rm tree}=64 \kappa^2  \cR^4$, and recalling that
in dimensional regularization, a polar term at $h$ loop comes with a coefficient of
 the form $\frac{L^{h\epsilon}}{\epsilon}=\frac{1}{\epsilon} + h \log L + \dots$ so a pole $1/\epsilon$ corresponds to a 
logarithmic term $h \log L$, 
we get
\be
\label{div2sugmul}
\cI_2^{D=9-2\epsilon} \sim -\frac{52\pi^4}{567} \log L\, \frac{\sigma_2^2}{2}\, \cR^4\ .
\ee
This is in precise agreement with \eqref{div1}.

\item In $D=8$, we get, from \eqref{E00L1} and \eqref{E01L2},
\be
\label{div2}
\begin{array}{lll}
\cE_{(0,0)}^{(2)}(\varphi,L) &\sim& 2\pi\log L \ ,\quad \\
\cE_{(0,1)}^{(2)}(\varphi,L) &\sim&   \frac{\pi}{3} \cE_{(0,0)}^{(2)}  \,
\log L  
+ \frac{2\pi^2}{3} (\log L )^2 \sim
 \frac{\pi}{3} \cE_{(0,0)}^{(2)}(\varphi,L) \, \log L
\end{array}
\ee
The first divergence, proportional to $\cR^4$, is simply the one-loop divergence in supergravity \cite{Green:1982sw}.
The second divergence, proportional to $\nabla^6\cR^4$,  comes both from the one-loop divergence of the $\cR^4$ form factor \eqref{div1r4},  and from the two-loop divergence \eqref{div2sug} in supergravity. The first term is guaranteed to be correctly normalized, since it contains the one-loop subdivergence of the genus-two amplitude, which is proportional to the
the $\cR^4$ coupling.
To check the normalization of the  two-loop divergence, 
we quote again the result in  \cite[(4.4)]{Bern:1998ug},
\be
\label{div28}
\cI^{D=8-2\epsilon}_{2,\rm subtracted} \sim  \left(\frac{\kappa}{2}\right)^6\,  \frac{1}{2(4\pi)^8} \left[ -\frac{1}{24\eps^2}+\frac{1}{144\epsilon}\right]\,\frac{(stu)^2}{3}\, M_4^{\rm tree}
\sim
 -\frac{2\pi^2}{3} (\log L)^2 \, 
\sigma_3\,\cR^4\  ,
\ee
where we identified $1/\epsilon^2 \sim 1/(2\epsilon^2)$, consistently with $
\frac{L^{h\epsilon}}{\epsilon^2}=\frac{1}{\epsilon^2} + h\frac{ \log L}{\epsilon}+\frac12 h^2 (\log L)^2+\dots$.
Comparing with the last term in \eqref{div2}, we see precise agreement, up to an overall sign.
This  sign difference
is in fact expected, since the result \eqref{div28} refers to the subtracted amplitude, where the
one-loop subdivergence has been cancelled against the triangle diagram with a one-loop
counterterm. As explained in \cite{Green:2010sp}, the relative normalisation of the sunset
and triangle diagrams is fixed by demanding that the coefficient of the $1/\epsilon$ pole in 
dimensional regularisation is analytic in the Mandelstam variables, and leads to a sign flip
in the coefficient of the $1/\epsilon^2$ pole.

\item In $D=7$, we get, from \eqref{E10L2},
\be
\label{div3}
\cE_{(1,0)}^{(3)}(\varphi,L) \sim \frac{4\pi^2}{3} \log L 
\ee
This should come from the two-loop divergence \eqref{div27} in supergravity. To check
the numerical factor, we quote the result in \cite{Bern:1998ug},
\be
\label{div27}
\cI_2^{D=7-2\epsilon} \sim \left(\frac{\kappa}{2}\right)^6\,  \frac{\pi}{3}\,  \frac{s^2+t^2+u^2}{2\epsilon(4\pi)^7}\, stu M_4^{\rm tree} = \frac{4\pi^2}{3} \log L\, \sigma_2\,\cR^4\ ,
\ee
in perfect agreement with \eqref{div3}. In particular, the factor $\pi/3$ in \eqref{div27} is identified
as the volume of the fundamental domain $\cF$.

\item In $D=6$, we get, from \eqref{E20L1}, \eqref{E10L2}, \eqref{E20L2}, \eqref{E01L3},
\be
\label{div4}
\begin{array}{lll}
\cE_{(1,0)}^{(4)}(\varphi,L) &\sim&\cE_{(0,0)}^{(4)}\,\log L \ ,\quad\\
\cE_{(0,1)}^{(4)}(\varphi,L) &\sim&5\zeta(3) \log L \\
\cE_{(2,0)}^{(4)}(\varphi,L) &\sim& 
\frac{5}{3}  \cE_{(1,0)}^{(4)}\, \log L   + \frac{7}{6} \cE_{(0,0)}^{(4)}\, (\log L )^2
\end{array}
\ee
The  first divergence, proportional to $\nabla^4\cR^4$, comes from the one-loop divergence of the $\cR^4$ form factor \eqref{div1r4} (which includes a contribution from  the one-loop subdivergence of the genus-two amplitude).
The divergence on the third line, proportional to $\nabla^8\cR^4$, may be rewritten as 
\be
\label{div44}
\cE_{(2,0)}^{(4)}(\varphi,L) 
\sim \frac{5}{3}  \cE_{(1,0)}^{(4)}(\varphi,L)\, \log L - \frac{1}{2} \cE_{(0,0)}^{(4)}\, (\log L )^2\ ,
\ee
exhibiting a contribution from the one-loop divergence of the $\nabla^4\cR^4$ form factor \eqref{div1d4r4}, and   from the two-loop divergence of the $\cR^4$ form factor 
\eqref{div2r4}. Note that \eqref{div44} predicts a divergence proportional to 
$\cE_{(1,0)}^{(4,2)}(\varphi) \log L$, which should arise from a  one-loop subdivergence in
the genus three contribution to the $\nabla^8\cR^4$ coupling \eqref{E20L3}.
Finally, the divergence  on the second line of \eqref{div4}, proportional to $\nabla^6\cR^4$, should come from the three-loop divergence \eqref{div3sug}.
To check the numerical coefficient, we quote the result in  \cite[(5.12)]{Bern:2008pv},
\be
\label{div3sug}
\cI_3^{D=6-2\epsilon} = \frac{5\zeta(3)}{(4\pi)^9 \epsilon} \left(\frac{\kappa}{2}\right)^8\, 
 (stu)^2\, M_4^{\rm tree} \sim 
\, 5\zeta(3)\, \log L\,  \sigma_3 \, \cR^4 \ ,
\ee
where we identified $1/\epsilon \sim 3 \log L$ as appropriate at 3-loop order.
This is in precise agreement with the coefficient of $\log L$ in \eqref{div4}.

\item In $D=5$, we get, from \eqref{E01L2},
\be
\label{div5}
\cE_{(0,1)}^{(5)}(\varphi,L) \sim \frac{5}{3} \cE_{(0,0)}^{(5)}\, \log L 
\ee
This comes from the two-loop divergence of the $\cR^4$ form factor \eqref{div2r4}, which is correctly
normalized since it includes
a contribution from the  two-loop subdivergence of the genus-three amplitude \eqref{E01L3}.

\item In $D=4$, we get, from \eqref{E01L1}, \eqref{E20L1}, \eqref{E01L2},\eqref{E20L2},
\be
\label{div6}
\begin{array}{lll}
\cE_{(0,1)}^{(6)}(\varphi,L) &\sim&\frac{5}{2\pi} \cE_{(1,0)}^{(6)}\,\log L \ ,\quad \\
\cE_{(2,0)}^{(6)}(\varphi,L) &\sim&
\frac{1}{4\pi} \left( 3  \cE_{(0,1)}^{(6)}
+ \frac12  (\cE_{(0,0)}^{(6)})^2 \right) \log L +  \frac{3}{2\pi^2} \cE_{(1,0)}^{(6)}\,(\log L )^2 
\end{array}
\ee
Ignoring possible contamination from infrared effects, 
the first divergence, proportional to $\nabla^6\cR^4$, should come from the one-loop divergence of the $\nabla^4\cR^4$ form factor \eqref{div1d4r4}  (which is correctly normalized since it includes
a contribution from the one-loop subdivergence of the genus-three amplitude \eqref{E01L3}).
The second divergence, proportional to $\nabla^8\cR^4$, may be rewritten as 
\be
\label{div66}
\cE_{(2,0)}^{(6)}(\varphi,L) &\sim&
\tfrac{1}{8\pi} (\cE_{(0,0)}^{(6)})^2 \, \log L 
+\tfrac{3}{4\pi} \cE_{(0,1)}^{(6)}(\varphi,L)\, \log L 
- \tfrac{3}{8\pi^2} \cE_{(1,0)}^{(6)}\,(\log L )^2 \ ,
\ee
exhibiting the  one-loop divergence of the $\nabla^6\cR^4$ and $(\cR^4)^2$ form factors 
\eqref{div1d6r4}, along with the  two-loop divergence of the 
$\nabla^4\cR^4$ form factor \eqref{div2d4r4}. Note that \eqref{div66} predicts divergences 
of the form $\cE_{(0,1)}^{(6,2)} (\log L)$, $\cE_{(1,0)}^{(6,1)} (\log L)^2$, which 
should arise from subdivergences of the genus-three amplitude \eqref{E20L3}.
Similarly, the term $\cE_{(1,0)}^{(6,2)} (\log L)^2$ should arise from a two-loop subdivergence of 
the genus-four string amplitude
at order $\nabla^8 \cR^4$, which is currently out of reach.

\item In $D=3$, we get, from \eqref{E20L2},
\be
\label{div7}
\cE_{(2,0)}^{(7)}(\varphi,L) \sim  \frac{2\beta \zeta(3)^2}{\pi^3 } \log L 
\ee
Barring possible effects from infrared divergences, this should come from the two-loop divergence of the $\nabla^6\cR^4$ and $(\cR^4)^2$ form factors \eqref{div2d6r4}. Unfortunately, we do not know how to derive
this coefficient independently.

\end{itemize}

\section{Discussion}
\label{sec_discuss}

In this note, we studied effective interactions of the form $\cE_{(p,q)}^{(d)}(\varphi,L)
\,\nabla^{4p+6q}\cR^4$ in the Wilsonian effective action $\cS(\Lambda)$ describing interactions
of massless particles at low energy in type II strings compactified on $T^d$. Perturbatively, these Wilsonian couplings are defined by 
restricting the integration over the moduli space $\cM_{h,n}$ of Riemann surfaces to the subset where the
proper time associated to any handle
is bounded by $L=1/(\alpha' \Lambda^2)$. Using recently available information on the
asymptotics of the integrands near boundaries of moduli space at genus one \cite{D'Hoker:2015foa}
and two \cite{Pioline:2015qha,DHoker:2017pvk,DHoker:2018mys}, we 
computed the large $L$ behavior of the coefficients $\cE_{(p,q)}^{(d)}(\varphi,L)$, and matched 
these terms against the UV divergences arising in maximal supergravity with a UV cut-off 
$\Lambda$, supplemented by higher-derivative stringy interactions. Since supergravity amplitudes 
are more easily computed in dimensional regularisation, 
we focused on logarithms arising in
various dimensions, at various order in the low energy expansion, although one could in principle also 
try to match power divergences in presence of a hard cut-off.

This matching
gives a strong consistency check on the structure of the asymptotic expansion of the 
integrands  $\cB^{(h)}_{(p,q)}$ near boundaries of moduli space, in particular on the powers of the degeneration parameter $t$ or $V$ appearing in the Laurent expansion, and it elucidates
the origin of the transcendental
coefficients which appear in these expansions. 
It follows from our analysis that the genus-one modular graph functions appearing in the expansion \eqref{nonkz}
at large $t$ are directly related to the genus-one integrands \eqref{Jpq}, while  the coefficients 
$\zeta(3), \zeta(5), \zeta(3)^2$ in the expansion \eqref{tropkz} in the
tropical limit $V\to 0$ are directly related to the tree-level interactions \eqref{Epqtree}.
In particular, this rules out a term proportional to $\zeta(7)$, which appears in intermediate
computations in \cite{DHoker:2018mys} but cancels in the final result. 
Combined with U-duality, this structure also
constrains the degenerations of the integrands at higher genus, even though the full integrand
is yet to be computed beyond genus two.  At the non-perturbative level,  U-duality
implies  that instanton corrections must also contribute divergent terms, although  the  Wilsonian effective action remains
to be defined beyond perturbation theory.

\begin{figure}[h]
\begin{center}
\begin{tikzpicture}[scale=.25]
\begin{scope}[xshift= -25  cm,yshift=0cm]
\draw [ultra thick] (-2.7,2.0) -- (2.7,2.0);
\draw [ultra thick] (-2.7,-2.0) -- (2.7,-2.0);
\draw (-3.0 ,0.0) circle (1.7 and 3.5);
\draw (3.0 ,0.0) circle (1.7 and 3.5);
\draw [ultra thick] (-3.7,-2.4) -- (-7,-4);
\draw [ultra thick] (-3.7,2.4) -- (-7,4);
\draw [ultra thick] (3.7,-2.4) -- (7,-4);
\draw [ultra thick] (3.7,2.4) -- (7,4);
\draw (0, -7.0) node{$(a)$};
\end{scope}
\begin{scope}[xshift= 0  cm,yshift=0.0cm]
\draw [ultra thick] (-2.7,2.0) -- (2.7,2.0);
\draw [ultra thick] (-2.7,-2.0) -- (2.7,-2.0);
\draw [ultra thick] (4,0) -- (10,-4);
\draw [ultra thick] (4,0) -- (10,4);
\filldraw [white]  (4,0) ellipse (2.5 and 2.5);
\draw  (4,0) ellipse (3.5 and 3.5);
\draw (6.6,0.5)  arc (-30 : -150:3);
\draw (5.6,-0.2)  arc (25 : 155:1.8);
\filldraw[white] (-3.0 ,0.0) circle (2.0 and 2.0);
\draw (-3.0 ,0.0) circle (1.7 and 3.5);
\draw [ultra thick] (-3.7,-2.4) -- (-7,-4);
\draw [ultra thick] (-3.7,2.4) -- (-7,4);
\draw (0, -7.0) node{$(b)$};
\end{scope}
\begin{scope}  [xshift= 25  cm,yshift=0.0cm]
\draw (02.1 ,0.0) circle (1.0 and 3.5);
\draw (-02.1 ,0.0) circle (1.0 and 3.5);
\draw [ultra thick] (-2.25 ,01.8) -- (-6, 3) ;
\draw [ultra thick] (-2.25 , -01.8) -- (-6, -3) ;
\draw [ultra thick] (2.25 , 01.8) -- (6, 3) ;
\draw [ultra thick] (2.25 , -01.8) -- (6, -3) ;
\draw [ultra thick] (-2.1 , -3) -- (2.1, -3) ;
\draw [ultra thick] (-2.1 , -0.) -- (2.1, -0.) ;
\draw [ultra thick] (-2.1 , 3) -- (2.1 , 3) ;
\draw [ultra thick] (0, -6.6) node{$(c)$};
\end{scope}
\end{tikzpicture}
\end{center}
\caption{(a)  Discontinuity of the one-loop amplitude. (b) Two-particle discontinuity of the two-loop amplitude.   (c) Three-particle discontinuity. 
}
\label{fig:disc}
\end{figure}
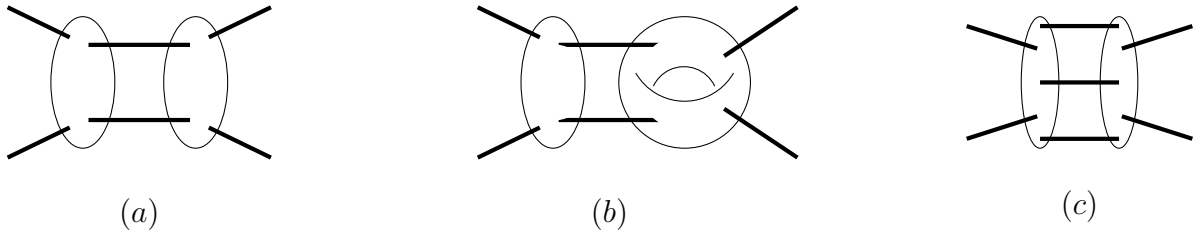

While the Wilsonian couplings depend on the cut-off $L$ and have the usual field redefinition
ambiguities, the full on-shell scattering amplitude \eqref{Adecomp} is  unambiguous and 
independent of $L$ (though it would depend on the IR cut-off  in $D\leq 4$). In dimensions where logarithms are present, this requires
a cancellation between analytic terms of the form $\log L \, \sigma_2^p \sigma_3^q\,\cR^4$ coming from the 
second term in \eqref{Adecomp}, and non-analytic contributions of the form 
$-\log(-s L \ell_s^2) \sigma_2^p \sigma_3^q \cR^4$ from the first term, leaving a term of the form 
$-\log (-s \ell_s^2) \, \sigma_2^p \sigma_3^q \cR^4$ in the amplitude. The latter are in turn computable by unitarity, since
they contribute to the imaginary part of the forward amplitude. This mechanism was studied in detail 
at genus-one in \cite{Green:2008uj}, where unitarity was used to relate the discontinuity of the one-loop
amplitude to the square of the four-point tree-level amplitude (diagram $(a)$ in Figure \ref{fig:disc}).
 
The same unitarity argument can in principle be applied at genus-two, but is considerably more involved. 
Indeed, one has to include both the
diagram $(b)$ with two-particle intermediate states, involving the product of a tree-level and
one-loop four-point amplitudes, and the diagram $(c)$ with three-particle intermediate states, involving
the product of two tree-level five-point amplitudes. In fact, rather than working order by order in the genus expansion,
it is more expedient to work at the non-perturbative level and express the imaginary part of the forward
amplitude as a sum of squares of on-shell $N+2$-point functions, where $N$ is the number of intermediate
states (for $N=2$ and $N=3$, this corresponds to the diagrams $(a)$ and $(c)$, where the oval now stands
for the full on-shell $N+2$-point function). 

Approximating the 4-point function in diagram $(a)$ and 5-point function in diagram $(b)$ by tree-level supergravity vertices, the sum over intermediate states and integral over the corresponding phase space  produces the discontinuity of the corresponding one-loop and two-loop
supergravity diagrams, from which one can read off the analytic structure of the amplitude. 
At one and two loops, the latter is 
schematically of the form \cite{Bern:1998ug}
\be
s^{\frac{D-8}{2}} \log(- s)\, \cR^4 + s^{D-7} \log(- s)\, \nabla^4 \cR^4 
\ee
These terms are indeed consistent\footnote{In $D=8$, there is also a term proportional to $[\log(- s)]^2 \nabla^6 \cR^4$, which is necessary to cancel $(\log L)^2 \nabla^6 \cR^4$ from the Wilsonian
coupling. In this sketchy discussion
we ignore such effects,  related to one-loop subdivergences of the two-loop amplitude.}
 with the $\log L$ terms predicted by \eqref{div1sug} and \eqref{div2sug}
in any dimension $D$. Including higher-derivative corrections of the form 
$\cE_{(p,q)}^{(d)} \nabla^{4p+6q}\cR^4$ to the on-shell quartic vertices  in diagram $(a)$, 
one obtains  a series
of corrections to the discontinuity, and therefore to the amplitude itself, of the form
\be
\cE_{(0,0)}^{(d)} s^{\frac{D-2}{2}} \log(- s)\, \cR^4 +
\cE_{(1,0)}^{(d)} s^{\frac{D+2}{2}} \log(- s)\, \cR^4 +
\left[ \cE_{(0,1)}^{(d)} + (\cE_{(0,0)}^{(d)})^2 \right] s^{\frac{D+4}{2}} \log(- s)\,  \cR^4 \nn\\ + \dots 
\ee
These terms agree with the $\log L$ terms from \eqref{div1r4}, \eqref{div1d4r4}, \eqref{div1d6r4}, etc,
justifying the use of on-shell vertices in these computations.  
In addition, there are also higher-derivative corrections to the on-shell five-point functions in 
diagram $(b)$, of the form $\nabla^{2n} \cR^5$ with $n\geq 1$ \cite{Stieberger:2009rr,Schlotterer:2012ny}. A subset 
of those corrections are  related by supersymmetry to the $\nabla^{4p+6q}\cR^4$ quartic couplings 
 with
$2n=4p+6q-2$ \cite{Green:2013bza}.
For those couplings, which are essentially obtained by connecting a quartic and cubic vertex using
a propagator, we expect contributions of the form
\be
\cE_{(0,0)}^{(d)} s^{D-4} \log(- s)\, \nabla^4 \cR^4 +
\cE_{(1,0)}^{(d)} s^{D-2} \log(- s)\,  \nabla^4 \cR^4 +
\left[ \cE_{(0,1)}^{(d)} + (\cE_{(0,0)}^{(d)})^2 \right] s^{D-1} \log(- s)\,  \nabla^4  \cR^4  \nn\\ \hspace*{-1cm} + \dots \qquad
\ee 
in agreement with the results \eqref{div2r4}, \eqref{div2d4r4}, \eqref{div2d6r4} of the naive
dimensional analysis. However, there are also
corrections to the  five-point function which are not related to  four-point vertices by 
supersymmetry \cite{Green:2013bza}. The leading such interaction, of order 
$\nabla^6 \cR^5$, will produce a term of the form $s^D  \log(- s)\, \nabla^4 \cR^4$, 
which predicts logarithmic divergences at order $\nabla^{12}\cR^4$ in $D=4$, far beyond
the order considered in this paper.  

As pointed out in \cite{Green:2010sp}, upon converting the non-analytic term $-\log (-s \ell_s^2) \, \sigma_2^p \sigma_3^q \cR^4$ 
in the amplitude from string frame to Einstein frame, using $\ell_s=\ell_D g_D^{2/(2-D)}$,  a term
of the form $\alpha \log L$ in the Wilsonian coupling $\cE_{(p,q)}^{(d)}(\varphi,L)$ implies a non-analytic term $\frac{4\alpha}{D-2} \log g_D$ in the weak coupling expansion of
the U-duality invariant coefficient $\cE_{(p,q)}^{(d)}(\varphi)$. Similarly, 
by the same logic as in the discussion of the $\nabla^6\cR^4$ interaction in $D=8$ \cite{Green:2010sp}, 
a term of the form $\tilde \alpha (\log L)^2$ in $\cE_{(p,q)}^{(d)}(\varphi,L)$ can be seen to imply a non-analytic term\footnote{In particular, the divergent terms in \eqref{div2} imply
$\cE_{(0,1)}^{(2),{\rm non-an.}}=-\frac{4\pi^2}{27}(\log g_8)^2 + \frac{2\pi}{9} ( \cE_{(0,0)}^{(2)} +
{\rm cte})\, \log g_8$, in agreement with \cite[(2.19)]{Pioline:2015yea}.}
 $-\frac12\tilde\alpha(\frac{4}{D-2} \log g_D)^2$. 
It follows
from the results in this note at order $\nabla^8\cR^4$  that
the U-duality invariant coefficient $\cE_{(2,0)}^{(d)}$ must contain the terms
\be
\label{E20nonan}
\begin{split}
\cE_{(2,0)}^{(d),{\rm non-an.}}=&
\left[ -\frac13 \cE_{(0,0)}^{(4)} \delta_{d,4} -\frac{81}{4\pi^2} \cE_{(1,0)}^{(6)} \delta_{d,6} 
\right] \, (\log g_D)^2\ \\
+&  
\left[
\frac{2\pi^2}{45}   \cE_{(0,0)}^{(0)} \, \delta_{d,0}
-\frac{416\pi^4}{3969} \delta_{d,1} + \frac{5}{3}  \left( \cE_{(1,0)}^{(4)} + \gamma\, 
\cE_{(0,0)}^{(4)}\right)\, \delta_{d,4} 
\right.\\&\left. 
+\frac{1}{2\pi} \left( 3  \cE_{(0,1)}^{(6)} + \eta\,  \cE_{(1,0)}^{(6)}
+ \frac12  (\cE_{(0,0)}^{(6)})^2 \right)  \delta_{d,6} 
+ \frac{8\beta \zeta(3)^2}{\pi^3 }\delta_{d,3} +\dots \right]\,
\log g_D 
\end{split}
\ee
generalizing similar results at order $\cR^4, \nabla^4\cR^4$ and $\nabla^6\cR^4$ 
\cite{Green:2010sp,Pioline:2015yea,Basu:2016kon}. In \eqref{E20nonan}, 
the dots indicate 
contributions which could in principle arise  at genus 3 or higher, and the 
coefficients $\gamma,\eta$ cannot be determined from  
this sketchy argument.

Thus, we see that the structure of the logarithms in the coefficients of the Wilsonian effective interactions is closely reflected 
by the analyticity properties of the scattering amplitude in a low energy expansion, and 
by the non-analytic terms in the  weak coupling expansion of the 
U-duality invariant coefficients. 
A more detailed discussion of these connections,
including effects of subdivergences  at genus two and beyond, is left for future work.

\medskip

\subsection*{Acknowledgments}
I wish to thank Eric d'Hoker and Michael B. Green for collaboration on  \cite{DHoker:2018mys} 
where  the problem addressed in this note was raised, and
for collaborating  at an initial stage on this follow-up project.
I am also grateful to Ashoke Sen, Guillaume Bossard, Henrik Johansson and Roji Pius 
for helpful discussions.
My research of is supported in part  by French state funds managed by  ANR in the context 
of the LABEX ILP (ANR-11-IDEX-0004-02, ANR-10-LABX-63).


\medskip

\appendix


\section{Constraints on $\nabla^{10} \cR^4$ interactions \label{sec_d10r4}}
In this section, we briefly discuss the implications of UV divergences in supergravity for
the string integrands computing $\nabla^{10} \cR^4$ interactions in the low energy
effective action.  At genus one, the integrand was computed in \cite[(7.28)]{D'Hoker:2015foa},
\be
 \cB^{(1)}_{(1,1)}=\frac{5}{6!} \left(336 C_{3,1,1} + 240 E_2 E_3 + 48 \zeta(3) E_2  - 
    \frac{1632}{5} E_5 + \frac{144}{5} \zeta(5) \right)\ ,
\ee
where $C_{3,1,1}$ is one of 
the  modular graph functions defined in \cite{D'Hoker:2015foa}. Using the asymptotics
derived in \cite[(6.2)]{D'Hoker:2015foa}, we have\be
\cB^{(1)}_{(1,1)} =\frac{4 \pi ^5 \tau_2^5}{66825}+\frac{\pi ^2 \tau_2^2 \zeta (3)}{63} 
+\frac{29 \zeta   (5)}{135}+\frac{\zeta (3)^2}{3 \pi  \tau_2}+\frac{49 \zeta (7)}{48 \pi ^2
   \tau_2^2}+\frac{\zeta (3) \zeta (5)}{12 \pi ^3 \tau_2^3} +
   \frac{21 \zeta (9)}{64
   \pi ^4 \tau_2^4}+\cO(e^{-2\pi\tau_2}) \nn\\
\ee
The resulting divergent terms in the truncated modular integral are given by \footnote{For $d=0$, this reduces to \cite[(3.22)]{Green:2008uj}, up to  normalisation. The finite term $\cE_{(1,1)}^{(0,1)}(\varphi)$ in $D=10/d=0$ is
computed in \cite[\S 7.4]{D'Hoker:2015foa}.}
\bea
\cE_{(1,1)}^{(d,1)}(\varphi,L) &= & \frac{8\pi^6}{66825} \frac{L^{\frac{d+8}{2}}}{\frac{d+8}{2}} 
+\frac{2\pi^3 \zeta(3)}{63} \frac{L^{\frac{d+2}{2}}}{\frac{d+2}{2}} 
+\frac{58 \pi \zeta(5)}{135}\frac{L^{\frac{d-2}{2}}}{\frac{d-2}{2}} 
+\frac{2\zeta(3)^2}{3}\, \frac{L^{\frac{d-4}{2}}}{\frac{d-4}{2}}  \nn \\
&&+\frac{49\zeta(7) }{24\pi}\frac{L^{\frac{d-6}{2}}}{\frac{d-6}{2}} 
+\frac{\zeta(3)\zeta(5)}{6\pi^2} \frac{L^{\frac{d-8}{2}}}{\frac{d-8}{2}} 
+\frac{21\zeta(9)}{32\pi^3} \frac{L^{\frac{d-10}{2}}}{\frac{d-10}{2}}
+\cE_{(1,1)}^{(d,1)}(\varphi)
\label{E11L1}
\eea
At genus two, the integrand $\cB^{(2)}_{(1,1)}$ is a modular graph function of weight 3, given by an integral of a sum of products
of three Arakelov-Green functions over the location of the four vertices \cite[(4.2)]{D'Hoker:2013eea}. 
The complete asymptotics of this modular graph function has not been fully analyzed yet, but its leading 
tropical limit  is known from the supergravity computation in \cite[(A.22)]{Green:2008bf}.\footnote{The
precise normalisation is given in \cite[(4.8)]{Bossard:2015oxa}. Note that these authors integrate over
a six-fold cover of the domain $\cF$, with no ordering on $L_1, L_2, L_3$, with 
measure $\de^3\Omega=\de L_1 \de L_2 \de L_3=V^2 \de V\de\mu(S)$.} 
Expressing
the result in terms of the modular local Laurent polynomials defined in Appendix \ref{sec_local}, we get
\bea
\cB^{(2)}_{(1,1)}&=& \frac{4\pi^3}{405V^3}
 \left[ -\frac{35}{66} A_{0,1} 
+ \frac{ 679}{429} A_{0,3}-\frac{2238}{143} A_{1,0}
+ \frac{147}{13} A_{1,2} + \frac{266}{17} A_{2,1} + \frac{145}{14} A_{3,0}  \right]
\nn\\
 &&  + \zeta(3) \tilde\cB_{(1,1)}^{(0)}
  + \frac{\zeta(5)}{\pi^2} \, V^2 \tilde\cB_{(1,1)}^{(2)}
   + \frac{\zeta(3)^2}{\pi^3} \, V^3   \tilde\cB_{(1,1)}^{(3)} 
  + \frac{\zeta(7)}{\pi^4} \, V^4   \tilde\cB_{(1,1)}^{(4)} 
  \nn\\
  &&
  + \frac{\zeta(3)\zeta(5)}{\pi^5} \, V^5   \tilde\cB_{(1,1)}^{(5)}   
    +  \, \frac{V^6}{\pi^6}   \left[ \zeta(9)\, \tilde\cB_{(1,1)}^{(6)}  + \zeta(3)^3\,
    \tilde\cB_{(1,1)}^{(6')} \right]
   + \cO(e^{-1/\sqrt{V}})\ .
\eea
The coefficients of the subleading terms on the second line are not known,
but they are expected to be linear combinations of the functions $A_{i,j}$. 
Rewriting this result in the variables $\tau_2=1/(V S_2)$, $t=S_2/V$ and 
reexpanding as $t\to \infty$, we find that the leading terms can be expressed 
in terms of genus-one modular graph functions as follows,
\be
 \cB^{(2)}_{(1,1)}(\Omega) &= & 
 \frac{2\pi^3  t^3 }{63} + \frac{4\pi^2t^2}{27} g_1(\tau,v) + \frac{\pi t}{270} 
\left( 157 E_2(\tau) + 75 g_1(\tau,v)^2 + 68 g_2(\tau,v) \right)  \nn\\
&& + \sum_{n=0}^3  \hat\cB^{(n)}_{(1,1)}(\tau,v)\,  t^{-n}    + \cO(e^{-2\pi t})
\ee
Here, the coefficients $\hat\cB^{(n)}_{(1,1)}(\tau,v)$ are modular graph functions of weight $n+3$.
Extracting the zero-th Fourier-Jacobi coefficient (see footnote \ref{foozero}), we get 
\be
 \int_{[-\frac12,\frac12]^3} \de u_1\, \de u_2 \, \de \sigma_1\, \cB^{(2)}_{(1,1)}(\Omega) &=& 
\frac{2\pi^3}{63} t^3  + \frac{116\pi t}{135}\, E_2   + \dots
\ee
The divergent terms in the truncated genus two integral are therefore
\be
 \label{E11L2}
\cE_{(1,1)}^{(d,2)}(\varphi,L) &=&  \frac{8\pi^4}{405}
\frac{L^{d}}{d}
 \left[
  -\tfrac{35}{66} F_{01}
+ \tfrac{679}{429} F_{03}-\tfrac{2238}{143} F_{10}
+ \tfrac{147}{13} F_{12} + \tfrac{266}{17} F_{21} + \tfrac{145}{14} F_{30}  \right](d)
\nn\\
 & +& 2\pi \zeta(3) \frac{\Lambda^{d-3}}{d-3} 
 \int_\cF \de\mu\, \tilde\cB_{(1,1)}^{(0)}
  + \frac{2\zeta(5)}{\pi} \, \frac{\Lambda^{d-5}}{d-5} 
 \int_\cF \de\mu\,  \tilde\cB_{(1,1)}^{(2)}
 \nn\\
 &
   +& \frac{2\zeta(3)^2}{\pi^2} \, \frac{\Lambda^{d-6}}{d-6} 
 \int_\cF \de\mu\, \tilde\cB_{(1,1)}^{(3)} 
  + \frac{2\zeta(7)}{\pi^3} \frac{\Lambda^{d-7}}{d-7} 
 \int_\cF \de\mu\, \tilde\cB_{(1,1)}^{(4)} + \dots
 \nn\\
 & +& \frac{\pi^3}{63}  \frac{L^{\frac{d+2}{2}}}{\frac{d+2}{2}} \cE_{(0,0)}^{(d,1)}(\varphi)
 + \frac{58\pi}{135}  \frac{L^{\frac{d-2}{2}}}{\frac{d-2}{2}} \cE_{(1,0)}^{(d,1)}(\varphi) 
  \nn\\ 
 &+& \frac{\pi}{2}  \frac{L^{\frac{d-4}{2}}}{\frac{d-4}{2}} \int_{\cF_1} \de\mu_1\, \hat \cB_{(1,1)}^{(0)}\Gamma_{d,d,1} (\varphi) 
  + \frac{\pi}{2}  \frac{L^{\frac{d-6}{2}}}{\frac{d-6}{2}} \int_{\cF_1} \de\mu_1\, \hat \cB_{(1,1)}^{(1)}\Gamma_{d,d,1} (\varphi) + \dots
\ee
where the dots denote additional $L$-dependent terms on which we have little control.

Using \eqref{Fijpole}, we find that 
the first term in \eqref{E11L2} leads to a logarithmic divergence in $D=10$, 
\be
\label{div011}
\cE_{(1,1)}^{(0,2)}(\varphi,L) \sim -    \frac{26\pi^4}{3645} \log L
\ee
which is in precise agreement 
with the two-loop divergence  computed in \cite{Bern:1998ug},
\be
\label{div210}
\cI_2^{D=10-2\epsilon} \sim -  \left(\frac{\kappa}{2}\right)^6\,  \frac{13}{25920}\, \frac{ (s^2+t^2+u^2) stu}{12\epsilon(4\pi)^{10}} \,\, stu\, M_4^{\rm tree}
\sim  -\frac{26\pi^4}{3645} \log L\, \sigma_2 \sigma_3 \, \cR^4\ .
\ee
In particular, the pole at $d=0$ is of first order, despite the fact that the various contributions
in the first line of \eqref{E11L2} are separately of order $1/d^2$ (see \eqref{Fijpole}). 
The structure
of the remaining terms on the second and third line of \eqref{div210} is consistent with the 
form factor divergences arising at two-loop in \eqref{div2r4}, \eqref{div2d4r4}, \eqref{div2d6r4}, 
and similarly for the $\nabla^8\cR^4$ form factor divergence proportional to $\zeta(7)$.
The last term on the fourth line, corresponding to the one-loop divergence  of the
$\nabla^4\cR^4$ form factor \eqref{div1d4r4} in $D=8$, nicely combines with the term
proportional to $\zeta(5)$ in \eqref{E11L1} so as to produce
\be
\cE_{(1,1)}^{(2)}(\varphi,L) \sim   \frac{58\pi}{135} \cE_{(1,0)}^{(2)}(\varphi)\,\log L\ ,
\ee
in turn predicting a one-loop subdivergence in the genus-three integrand at order $\nabla^{10}\cR^4$.
Requiring
that the coefficients of the logarithmic divergence in $D=4$ recombines into
the U-duality invariant coupling $\cE_{(2,0)}^{(6)}$, we predict\footnote{Here we ignore a possible contribution of the diagram $(c)$ in Figure \ref{fig:oneloop1}, involving a $\nabla^6\cR^5$ quintic coupling unrelated to $\nabla^8\cR^4$
by non-linear supersymmetry, of the type considered in   \cite{Green:2013bza}.}
\be
 \int_{[-\frac12,\frac12]^3} \de u_1\, \de u_2 \, \de \sigma_1\ 
 \hat \cB_{(1,1)}^{(1)} =\frac{49}{12\pi^2}  \cB_{(2,0)}^{(1)}  = 
  \frac{49}{144\pi^2}(D_4+9 E_2^2 + 6 E_4) 
\ee 
Similarly, requiring that the coefficients of the logarithmic divergence in $D=6$ recombines into
a linear combination of the U-duality invariant couplings $\cE_{(0,1)}^{(4)}$ and $\left[\cE_{(0,0)}^{(4)}\right]^2$ with rational coefficients, we see that the constant term of  $\hat \cB_{(1,1)}^{(0)}$ must be a linear combination
of $E_3$ and $\zeta(3)$. 

\section{Some properties of modular local polynomials}
\setcounter{equation}{0}
 \label{sec_local}
 
In the tropical limit $V\to 0$, the genus-two integrands have a finite Laurent series expansion
\be
\cB^{(2)}_{(p,q)}(\Omega) = \sum_{n=-(2p+3q-2)}^{4p+6q-4}  V^n \, \tilde\cB_{(p,q)}^{(n)}(\ttau) + \cO(e^{-1/\sqrt{V}})\
\ee
where $\tilde\cB_{(p,q)}^{(n)}(\ttau)$ are modular local Laurent polynomials, a class 
of non-holomorphic modular functions
first encountered in the study of two-loop supergravity amplitudes \cite{Green:2008bf} and
further developed in the mathematics literature \cite{zbMATH06251204,MR3554498}.
These functions are invariant under the action of $GL(2,\ZZ)$ by fractional linear transformations, 
\be
\label{GL2act}
\ttau\mapsto \begin{cases} \frac{a\ttau+b}{c\ttau+d} & ad-bc=1\\
\frac{a\bar\ttau+b}{c\bar\ttau+d} &ad-bc=-1
\end{cases}
\ee
but they are not everywhere smooth, hence not a modular function
in the usual sense. Their singular locus is given by the union of real-codimension one loci 
\be
\label{singloc}
a|\ttau|^2 + b \ttau_1 + c=0\ ,\qquad b^2-4ac=1\ ,
\ee
corresponding to the tropical limit of the separating degeneration divisor $v=0$ and its images under
$Sp(4,\ZZ)$. Inside the standard fundamental domain of $GL(2,\ZZ)$,
\be
\cF = \{ \ttau=\ttau_1+\I\ttau_2 \quad | \quad 0<\ttau_1<\frac12, \quad |\ttau|^2>1 \} = \cF_1/\ZZ_2 \ ,
\ee
they are Laurent polynomial in $\ttau_2$, with coefficients which are polynomial in $\ttau_1$, invariant
under $\ttau_1\to 1-\ttau_1$. 
A basis of such local modular functions is given by \cite{ZagierPC,DHoker:2018mys}
\be
\label{covddef}
A_{i,j} = D_{-2n}^{(n)} (u^i v^j)
\ee
where 
\be
u= \ttau ^2 (\ttau -1)^2 \ ,\quad v=\ttau ^2-\ttau +1\ ,\quad n=3i+j\ .
\ee  
The operator $D_{-2n}^{(n)}$ is the $n$-th iteration of the Maass raising operator acting on 
modular forms of weight $-2n$,  
\be
D_{-2n}^{(n)} = \frac{(-2i)^n n!}{(2n)!}\, 
 \sum_{m=0}^n \begin{pmatrix} n \cr m \end{pmatrix}\frac{(-n-m)_m}{(\ttau -\bar \ttau )^m}\, \frac{\partial^{n-m}}{\partial \ttau ^{n-m}}
 \ee
where $(k)_m=k(k+1)\dots (k+m-1)$ is the Pochhammer symbol.  In the standard fundamental domain, the modular function $A_{i,j}$ takes the form
\be
\label{Aijexp}
A_{i,j}(\ttau) =\sum_{k=0}^{2i+j} A_{i,j}^{(k)}(\ttau_1)\, \ttau_2^{i+j-2k}
\ee
where $A_{i,j}^{(k)}$ is a polynomial of degree $2k$ in $\ttau_1$, or rather a polynomial of degree $k$ in $T_1=\ttau_1(1-\ttau_1)$. 
Away from the singular loci \eqref{singloc}, $A_{i,j}$ is an eigenmode of the Laplace operator 
$ \Delta_\ttau =\ttau_2^2 (\partial_{\ttau_1}^2+\partial_{\ttau_2}^2)$ on the upper half-plane,
 \be
 \label{LapAij}
(\Delta_\ttau - n(n+1)) A_{i,j} =0\ ,
\ee
where $n=3i+j$.
On the locus $\ttau_1=0$, this equation acquires a source term proportional to 
$\delta(\ttau_1)$ \cite{Green:2008uj}.
In order of increasing $n$, the first few  $A_{i,j}$'s  are given explicitly in the
standard fundamental domain by 
 $A_{0,0}=1$ and
 \begin{subequations}
\bea
A_{0,1}&=& \ttau_2+\frac{1-T_1}{\ttau_2}\\
   A_{02} &=& \ttau_2^2+(1-2 T_1)+\frac{(T_1-1)^2}{\ttau_2^2}\\
A_{1,0}&=&\frac{\ttau_2}{5}+\frac{1-6
   T_1}{5\ttau_2}+\frac{T_1^2}{\ttau_2^3}\\
    A_{0,3}&=&  \ttau_2^3+\left(\frac{6}{5}-3 T_1\right) \ttau_2+\frac{3 \left(5 T_1^2-7
   T_1+2\right)}{5
   \ttau_2}-\frac{(T_1-1)^3}{\ttau_2^3}   \\
   A_{1,1} &=& \frac{\ttau_2^2}{7}+\left(\frac{12}{35}-\frac{9 T_1}{7}\right)+\frac{3 T_1 (5
   T_1-3)+1}{7 \ttau_2^2}-\frac{(T_1-1)
   T_1^2}{\ttau_2^4}\\
   A_{2,0} &=& \frac{\ttau_2^2}{33}+\left(\frac{10}{77}-\frac{20 T_1}{33}\right)+\frac{10 T_1
   (7 T_1-2)+1}{33 \ttau_2^2}+\frac{2 (3-14 T_1) T_1^2}{11
   \ttau_2^4}+\frac{T_1^4}{\ttau_2^6}\\
   A_{1,2} &=& \frac{\ttau_2^3}{9}+\left(\frac{8}{21}-\frac{4 T_1}{3}\right)
   \ttau_2+\frac{2 \left(35 T_1^2-25 T_1+4\right)}{21
   \ttau_2}+\frac{-28 T_1^3+36 T_1^2-12 T_1+1}{9
   \ttau_2^3} \nn \\ && +\frac{(T_1-1)^2
   T_1^2}{\ttau_2^5}
   \eea
   \bea
   A_{3,0} &=& \frac{\ttau_2^3}{221}-\frac{21 (22 T_1-5) \ttau_2}{2431}+\frac{105 \left(33
   T_1^2-12 T_1+1\right)}{2431 \ttau_2}+\frac{-924 T_1^3+378 T_1^2-42
   T_1+1}{221 \ttau_2^3}\nn\\
   &&+\frac{3 T_1^2 \left(33 T_1^2-12
   T_1+1\right)}{17 \ttau_2^5}-\frac{3 \left(T_1^4 (22 T_1-5)\right)}{17
   \ttau_2^7}+\frac{T_1^6}{\ttau_2^9}
   \eea
     \end{subequations}
It will be useful to note that except for $A_{00}$, all $A_{ij}$ vanish on the line $S_1=1/2, S_2>\sqrt{3}/2$. On the line $S_1=0, S_2>1$, they are continuous but not differentiable, with the left/right derivative
$P_{i,j}(\ttau) \equiv  \pm\partial_{\ttau_1}A_{ij}\vert_{\ttau_1=0^\pm}$ given by
\be
\begin{split}
P_{0,1}=&-\frac{1}{\ttau_2}\ ,\quad
P_{0,2}=-2\left(1+\frac{1}{\ttau_2^2}\right)\ ,\quad
P_{1,0}=-\frac{6}{5\ttau_2}\ , \\
P_{1,1}=&-\frac97\left(1+\frac{1}{\ttau_2^2}\right)\ ,\quad
P_{2,0}=-\frac{20}{33}\left(1+\frac{1}{\ttau_2^2}\right)
\\
P_{1,2}=&-\frac{4 \ttau_2}{3}-\frac{50}{21 \ttau_2}-\frac{4}{3 \ttau_2^3}\ ,\quad
P_{2,1}=-\frac{75 \ttau_2}{143}-\frac{490}{429 \ttau_2}-\frac{75}{143  \ttau_2^3}\ ,\quad
   \\
P_{0,3}=&-3 \ttau_2-\frac{21}{5  \ttau_2}-\frac{3}{\ttau_2^3}\ ,\quad
P_{3,0}=-\frac{42 \ttau_2}{221}-\frac{1260}{2431 \ttau_2}-\frac{42}{221 \ttau_2^3}\ .
\end{split}
\eea
It will also be useful to record the zero-mode with respect to $\ttau_1$ for $\ttau_2>1$,
\begin{subequations}
\label{LiIntA}
\bea
\int_0^{1/2} \de\ttau_1\, A_{0,1} &=& \frac{\ttau_2}{2}+\frac{5}{12\ttau_2} \\
\int_0^{1/2} \de\ttau_1\, A_{0,2} &=&\frac{\ttau_2^2}{2}+\frac13+\frac{7}{20\ttau_2^2} \\ 
\int_0^{1/2} \de\ttau_1\, A_{1,0} &=&\frac{\ttau_2}{10}+\frac{1}{60\ttau_2^3}\\
\int_0^{1/2} \de\ttau_1\, A_{0,3} &=&\frac{\ttau_2^3}{2}+\frac{7 \ttau_2}{20}+\frac{3}{10
   \ttau_2}+\frac{83}{280 \ttau_2^3}\\
\int_0^{1/2} \de\ttau_1\, A_{1,1} &=&\frac{\ttau_2^2}{14}+\frac{9}{140}+\frac{11}{840\ttau_2^4} \\
\int_0^{1/2} \de\ttau_1\, A_{2,0} &=&\frac{\ttau_2^2}{66}+\frac{10}{693}+\frac{11}{1260\ttau_2^6} \\
\int_0^{1/2} \de\ttau_1\, A_{1,2} &=&\frac{\ttau_2^3}{18}+\frac{5 \ttau_2}{63}+\frac{1}{21
   \ttau_2}+\frac{13}{1260 \ttau_2^5}\\
\int_0^{1/2} \de\ttau_1\, A_{2,1} &=&\frac{3 \ttau_2^3}{286}+\frac{35 \ttau_2}{1716}+\frac{25}{2574
   \ttau_2}+\frac{17}{27720  \ttau_2^7}\\
   \int_0^{1/2} \de\ttau_1\, A_{3,0} &=&\frac{\ttau_2^3}{442}+\frac{14 \ttau_2}{2431}+\frac{21}{9724
   \ttau_2}+\frac{1}{24024  \ttau_2^9} 
\eea
   \end{subequations}
Note that several of the coefficients $A_{i,j}^{(k)}(\ttau_1)$ with $0<k<2i+j$ integrate to zero.

In order to define the renormalized integral in \S\ref{sec_trunc}, we need to compute integrals of the form 
\be
\label{IntttauA}
F_{ij}(s) = \int_{\cF} \de\mu\, \ttau_2^{-s} \, A_{ij}(\ttau)\ ,\qquad \de\mu = 2 \frac{\de \ttau_1\de\ttau_2}{\ttau_2^2}.
\ee
Inserting the expansion \eqref{Aijexp}, we see that the integral converges for $\Re(s)>i+j-1$, and 
 that simple poles occur at $s=i+j-2k-1$ for $k=0,\dots 2i+j$, with residue proportional to $\int_0^{1/2}\de\ttau_1\,A_{i,j}^{(k)}(\ttau_1)$. Using \eqref{LiIntA} we deduce the pole structure for
low values of $(i,j)$:
\begin{itemize}
\item $F_{0,0}$ has simple poles at $s=-1$, with residue $1$;
\item $F_{0,1}$ has simple poles at $s\in \{0,-2\}$, with residues $\{1,\frac56\}$, respectively; 
\item $F_{0,2}$ has simple poles at $s\in\{1,-1,-3\}$, with residues $\{1, \frac23, \frac7{10}\}$; 
\item $F_{1,0}$ has simple poles at $s\in\{0,-4\}$, with residues $\{\frac15,\frac1{30}\}$; 
\item $F_{0,3}$ has simple poles at $s\in\{2,0,-2,-4\}$, with residue $\{1,\frac{7}{10},\frac35,\frac{83}{140}\}$; 
\item $F_{1,1}$ has simple poles at $s\in\{1,-1,-5\}$, with residues $\{\frac17,\frac9{70},\frac{11}{420}\}$; 
\item $F_{2,0}$ has simple poles at $s\in\{1,-1,-7\}$, with residues $\{\frac{1}{33},
\frac{20}{693},\frac{1}{630}\}$;
\item $F_{1,2}$ has simple poles at $s\in\{2,0,-2,-6\}$, with residues $\{\frac{1}{9},
\frac{10}{63},\frac{2}{21},\frac{5}{63}\}$;
\item $F_{2,1}$ has simple poles at $s\in\{2,0,-2,-8\}$, with residues $\{\frac{3}{143},
\frac{35}{858},\frac{25}{1287},\frac{17}{13860}\}$;
\item $F_{3,0}$ has simple poles at $s\in\{2,0,-2,-10\}$, with residues $\{\frac{1}{221},
\frac{28}{2431},\frac{21}{4862},\frac{1}{12012}\}$;
\end{itemize}
The full integral \eqref{Intttau} can be evaluated by integration by parts, 
using the fact \eqref{LapAij} that  $A_{i,j}$ is
an eigenmode of $\Delta_\ttau$ away from $\ttau_1=0$. Following the same steps as in  \cite[(2.45)]{Pioline:2015nfa}, we get, for $n=3i+j>0$,
\be
\label{intFs}
\begin{split}
\left[ 1- \frac{s(s+1))}{n(n+1)} \right]  \int_{\cF}\de\mu\, \ttau_2^{-s} A_{i,j}(\ttau) = \frac{1}{n(n+1)} \int_{\partial \cF} 
\left( \ttau_2^{-s}\,\star\,\de A_{i,j}-A_{i,j}\, \star\de\ttau_2^{-s} \right) \hspace*{2cm}\\
=  -\frac{2}{n(n+1)} \int_1^{\infty}\, P_{ij}(\ttau_2)\, \ttau_2^{-s} \de\ttau_2 
- \frac{2s}{n(n+1)} \int_0^{1/2} \de\ttau_1\, (1-\ttau_1^2)^{\frac{-s-1}{2}}\,
A_{ij}\left(\ttau_1,\sqrt{1-\ttau_1^2}\right)  
\end{split}
\ee
where the two terms on the last line correspond to the boundary contribution at $\ttau_1=0$ and at $|\ttau|^2=1$ (there is no contribution from the  boundary at $\ttau_1=1/2$ since both $A_{i,j}$ and
the normal derivative $\star\de A_{i,j}$ vanish there). The first integral is trivially evaluated, while the second 
integral is a linear combination of integrals of the form \cite[(2.37)]{Pioline:2015nfa}
\be
c_n(\gamma) :=\int_0^{1/2}\, \de x\, x^{2n} (1-x^2)^{-\gamma} = \frac{4^{-n-1}}{n+\frac12}
\, _1F_2\left(\gamma,n+\frac12;n+\frac32;\frac14\right)\ .
\ee
Poles in $F_{i,j}(s)$ a priori occur when $s=n, s=-n-1$ or when $ \int_1^{\infty}\, P_{ij}(\ttau_2)\, \ttau_2^{-s} \de\ttau_2$ has a pole, since the last term in \eqref{intFs} is smooth. 
In order for the apparent pole at $s=n$ to cancel,
the r.h.s. of \eqref{intFs} must vanish at that value. 
Using this we find 
\be
\label{Fijzero}
F_{0,0}(s)=\frac{\pi}{3} +\cO(s)\ ,\quad F_{0,2}, F_{1,1}, F_{2,0} = \cO(s)\ ,
\ee
while 
\be
\label{Fijpole}
\begin{split}
F_{0,1}(s)=&\frac{1}{s}+\frac12+\cO(s)\ ,\quad
F_{1,0}(s)=\frac{1}{5s}+\frac1{60}+\cO(s)
\\
F_{1,2}(s)=&\frac{10}{63s}+\frac{1}{189}+\cO(s), \quad F_{2,1}=\frac{35}{858s}+\frac{5}{6864}+\cO(s)\ ,\quad 
\\
F_{0,3}(s)=&\frac{7}{10s}+\frac{7}{120}+\cO(s)\ ,\quad
F_{3,0}(s)=\frac{28}{2431s}+\frac{14}{109395}+\cO(s)\ .
\end{split}
\ee
The finite terms in these expansions have been used in obtaining \eqref{div011}.


\providecommand{\href}[2]{#2}\begingroup\raggedright\endgroup

\end{document}